\documentclass[11pt]{iopart}
\expandafter\let\csname equation*\endcsname\relax     
\expandafter\let\csname endequation*\endcsname\relax  
\usepackage{amssymb}
\usepackage{amsmath}
\usepackage{bbm}
\usepackage{graphicx}
\usepackage{color}

\begin{document}

\title{Renormalization of lattice field theories with  infinite-range wavelets}
\author{Pascal Fries$^1$, Ignacio Reyes$^{1,2}$, Johanna Erdmenger$^1$, and\\ Haye Hinrichsen$^1$}
\address{$^1$ Faculty for Physics and Astronomy, University of W\"urzburg, Am Hubland\\ 97074  W\"urzburg, Germany}
\address{$^2$ Instituto de F\'isica, Pontificia Universidad Cat\'olica de Chile, \\
Casilla 306, Santiago, Chile}

\begin{abstract} 
We present a new exact renormalization approach for quantum lattice models leading to long-range interactions. The renormalization scheme is based on wavelets with an infinite support in such a way that the excitation spectrum at the fixed point coincides with the spectrum of the associated short-range continuum model in an energy range below an upper cutoff imposed by the lattice spacing. As a consequence, the conformal towers of spectrum are exactly realized on the lattice up to a certain energy scale. We exemplify our approach by applying it to free bosons and to free fermions in $1+1$ dimensions, as well as to the Ising model. The analysis is also motivated by  tensor network approaches to the AdS/CFT correspondence since our results may be useful for a qualitatively new construction of holographic duals complementary to previous approaches based on finite-range wavelets.
\end{abstract}

\vfill
Submitted to JSTAT: Special Issue in memory of Vladimir Rittenberg
\maketitle
\parskip 1mm

\def\comment#1{\color{red}[\textbf{#1}]\color{black}}
\def\mark#1{\color{red}#1 \color{black}}

\section{Introduction}

The concept of the renormalization group (RG) is one of the cornerstones in the theory of phase transitions and critical phenomena.  It is based on the idea that a scale transformation can effectively be absorbed in changing the parameters of the system. If these parameters flow to a non-trivial fixed point under renormalization, the corresponding system is expected to be scale-invariant, exhibiting interesting universal features at large scales.

Renormalization as a continuous change of scale is a particularly natural concept in the context of continuum models or field theories. For discretized theories and lattice models, however, RG-transformations are much harder to implement since scale changes have to comply with the underlying lattice structure.  In fact, apart from very few exceptions, most renormalization schemes on lattices are approximate in the sense that  contributions of higher order are neglected.  One of the few exact RG schemes is the  block-spin renormalization group of the one-dimensional Ising model~\cite{nelson1975soluble} on an infinite lattice, where a coarse-graining in pairs of spins is \textit{exactly} equivalent to a parameter change in the Hamiltonian.  Unfortunately, this renormalization scheme is no longer exact in higher dimensions.

The present study is motivated by recent results obtained by Evenbly and White~\cite{evenbly2016entanglement} who demonstrated that discrete wavelet transforms~\cite{chui2016introduction} can be used to establish a real-space RG transformation for quantum systems on a lattice in the context of free-particle systems. Using finite-range Daubechies wavelets~\cite{daubechies1988orthonormal} in a multiscale entanglement renormalization ansatz (MERA)~\cite{vidal2008class} they were able to approximate the ground state of the critical Ising model. The approximative nature of their approach can be traced back to the fact that the employed wavelet filter functions have a finite spatial range. Therefore, the question arises whether it is possible to improve the accuracy by increasing the range of the wavelets and whether we can reach a limit in which such a MERA-inspired RG scheme becomes exact in the sense that the lattice system at the RG fixed point behaves exactly in the same way as its continuum counterpart.

In this paper we demonstrate that such a limit does indeed exist, giving explicit examples of exact lattice RG schemes for simple integrable systems. The RG procedures used here are based on long-range wavelets in the limit of perfect momentum separation,  known in wavelet theory as the Shannon limit~\cite{cattani2008shannon}.  Roughly speaking, such transformations divide the momentum space of the theory into two equal halves separating low and high wave numbers.  Studying various types of free theories (free bosons, free Dirac fermions, and the Ising model), we find that such a transformation completely decouples the low- and high-momentum modes, providing a consistent exact renormalization scheme  on a lattice. The exactness, however, comes at the price that the RG flow generates algebraic long-range interactions. In many situations the emergence of such long-range interactions is undesirable. Here  however, we would like to adopt the opposite viewpoint, studying long-range interactions as an interesting subject on its own and taking advantage of them. The main results are:

\begin{itemize}
 \item The filter functions in the limit of perfect momentum separation are unique and lead to non-local interactions which decay algebraically (i.e. with a power law) with distance. 
 \item For free theories,  the RG scheme produces exactly the same energy excitation spectrum as in the continuum limit, bounded by a cut-off energy given by the inverse lattice spacing. 
 \item The RG fixed point exhibits the same universal properties as the corresponding short-range model at criticality.
 \item At the long-range RG fixed point we find that various physical properties, e.g. the specific heat, are much less affected by lattice artifacts than in the short-range case.
 \item In fermionic models, the problem of fermion doubling, which emerges as an artifact in fermion lattice models with short-range interactions, is no longer present since the doubled UV-modes are eliminated already after the first RG step.
\end{itemize}

Wavelets may also provide a new approach for constructing higher-dimensional hyperbolic spaces, which is of  relevance for the AdS/CFT correspondence. It is a well-known fact that within AdS/CFT, the renormalization scale can be viewed as an extra dimension. In the context of MERA networks, this was first pointed out in \cite{Swingle:2009bg}. More recently, the approach of exact holographic mappings \cite{Qi:2013caa} has been connected to free fermion renormalization via a Haar wavelet  \cite{Lee:2015vla} and also extended to more general wavelets with a larger but finite range in \cite{Singh:2016mxd,PhysRevB.96.245103}. While previous applications of wavelets to AdS/CFT-inspired lattice models used wavelets with compact support, the main ingredient of the present work is that  in the Shannon limit, the wavelets have an infinite range and are therefore highly non-local. Since our goal is to achieve an exact decoupling of the IR and UV modes at every renormalization step, the `bulk' correlation functions in the holographic direction vanish, and therefore the standard methods for defining an emergent metric in the wavelet approach  \cite{Qi:2013caa,PhysRevB.96.245103} are not directly applicable. However, as we shall see, our approach provides an exact realization of Verma modules up to a certain energy scale and therefore may provide a new alternative approach to constructing  gravity duals.

As a concrete application to CFTs, we shall illustrate the usefulness of the Shannon limit by examining the spectrum of the quantum Ising model on the lattice at criticality. Although conformal symmetry is broken by the lattice, the non-local wavelet flow allows to recover the correct Verma module spectrum up to a scale of the inverse lattice space. Although this is not a strongly coupled theory, it provides a first step towards the better understanding of scale invariant theories on the lattice. Indeed, the Ising model was studied in connection to holography \cite{Castro:2011zq}.

The paper is organized as follows. In the following section we first demonstrate the key concept in the example of a single free quantum particle. In the third section the long-range filter functions are introduced on infinite as well as finite lattices. In the sections 4-6 this concept is applied to free bosons, free fermions, and the critical Ising model. After some concluding remarks, technical details are given in an appendix.

\section{Free quantum particle}

As the simplest example that outlines the main idea let us consider a free quantum particle in 1+1 dimensions which evolves according to the non-relativistic Schr\"odinger equation (setting $\hbar=m=1$)
\begin{equation}
\label{schroedinger}
i \frac{\partial}{\partial t}\psi(x,t) \;=\; -\frac12 \nabla^2 \psi(x,t) \,.
\end{equation}
This equation is invariant under the scale transformation $x \to \lambda x, \,\, t \to \lambda^z t$  with the dynamical exponent $z=2$, which is reflected in the quadratic dispersion relation
\begin{equation}
\label{quadraticdispersion}
\omega = \frac12 k^2\,.
\end{equation}
Let us now consider a discretized variant of the Schr\"odinger equation on an infinite lattice with equidistant sites positioned at $x_n = na$,  where $a$ is the lattice spacing and $n \in \mathbb Z$. Here it is customary to discretize the Laplacian on the r.h.s. of Eq.~(\ref{schroedinger}) by
\begin{equation}
\label{discreteSchroedinger}
i \frac{\partial}{\partial t} \psi_n \;=\; -\frac1{2a^2} \bigl( \psi_{n-1}-2\psi_n+\psi_{n+1}\bigr)\,,
\end{equation}
where $\psi_n:=\psi(x_n,t)$. It is easy to verify that the discretized Laplacian leads to a modified dispersion relation of the form
\begin{equation}
\omega = 1-\cos(a k)\,, \qquad k \in [-\tfrac{\pi}{a},\tfrac{\pi}{a}]
\end{equation}
which reproduces Eq. (\ref{quadraticdispersion}) in the limit $|k| \to 0$ but deviates from the quadratic law for all $k\neq 0$ due to the finite lattice spacing.

Is it possible to find a modified Laplacian $\Delta^*$ that behaves on the lattice \textit{exactly} in the same way as in the continuum limit, exhibiting a clean quadratic dispersion? This is in fact an easy exercise. Because of linearity and translational invariance such a Laplacian has to be a convolution product of the form
\begin{equation}
(\Delta^* \psi)_n \;=\; \sum_{n'} \Delta^*_{n-n'} \,\psi_{n'}\,,
\end{equation}
with coefficients $\Delta^*_r$ given by the inverse Fourier transform of the dispersion relation:
\begin{equation}
\label{longrangelaplacian}
\Delta^*_r = -\frac{a}{2\pi} \int_{-\pi/a}^{\pi/a} k^2\,e^{i a k r}\, d k \;=\; 
\left\{
\begin{array}{ll} 
-\frac{\pi^2}{3a^2} & \textrm{ if } r=0\\ 
-\frac{2(-)^{r}}{r^2a^2} & \textrm{ otherwise }
\end{array}
\right. 
\end{equation}
This modified discrete Laplacian, known in lattice quantum field theory as the SLAC-derivative~\cite{quinn1986new}, is advantageous in many ways. For example, the solutions of the discretized Schr\"odinger equation $i \dot \psi_n = \frac12 (\Delta \psi)_n$ will coincide \textit{exactly} with those of the continuous Schr\"odinger equation for all initial conditions in the common functional space, i.e., with Fourier components confined to the Brillouin zone $|k|\leq \pi/a$.

\begin{figure}
\centering\includegraphics[width=130mm]{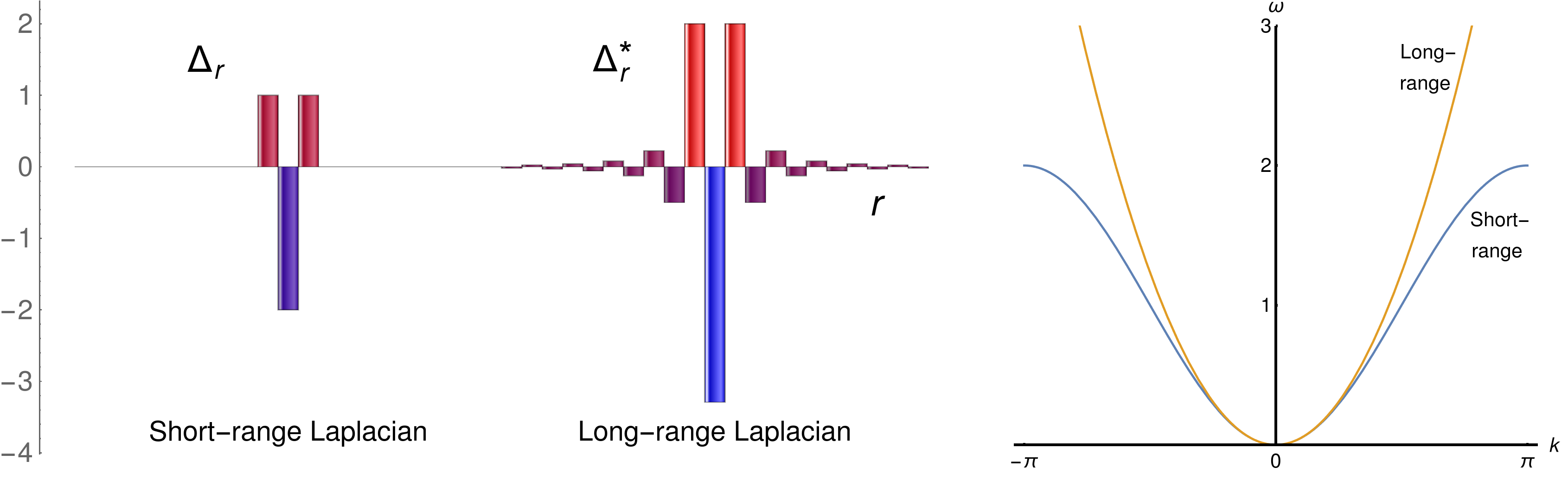}
\vspace{-2mm}
\caption{Left: Magnitude of the components of the short-range lattice Laplacian~$\Delta$ and its long-range counterpart $\Delta^*$ defined in  (\ref{longrangelaplacian}). Right: Corresponding dispersion relations.}
\label{fig:laplacian}
\end{figure}

However, the modified Laplacian \eqref{longrangelaplacian} is no longer local as \eqref{discreteSchroedinger}, but instead involves algebraically decaying long-range interactions with alternating signs (see Fig.~\ref{fig:laplacian}). Although this may seen as be a major drawback in various applications, the aim of the present work is to appreciate such long-range interactions and to investigate their properties. In particular, we are going to show that such interaction structures arise naturally as a fixed point under a certain kind of renormalization. 

\section{Wavelet-inspired renormalization scheme}
\label{sec:RGScheme}

\begin{figure}[b]
\centering\includegraphics[width=100mm]{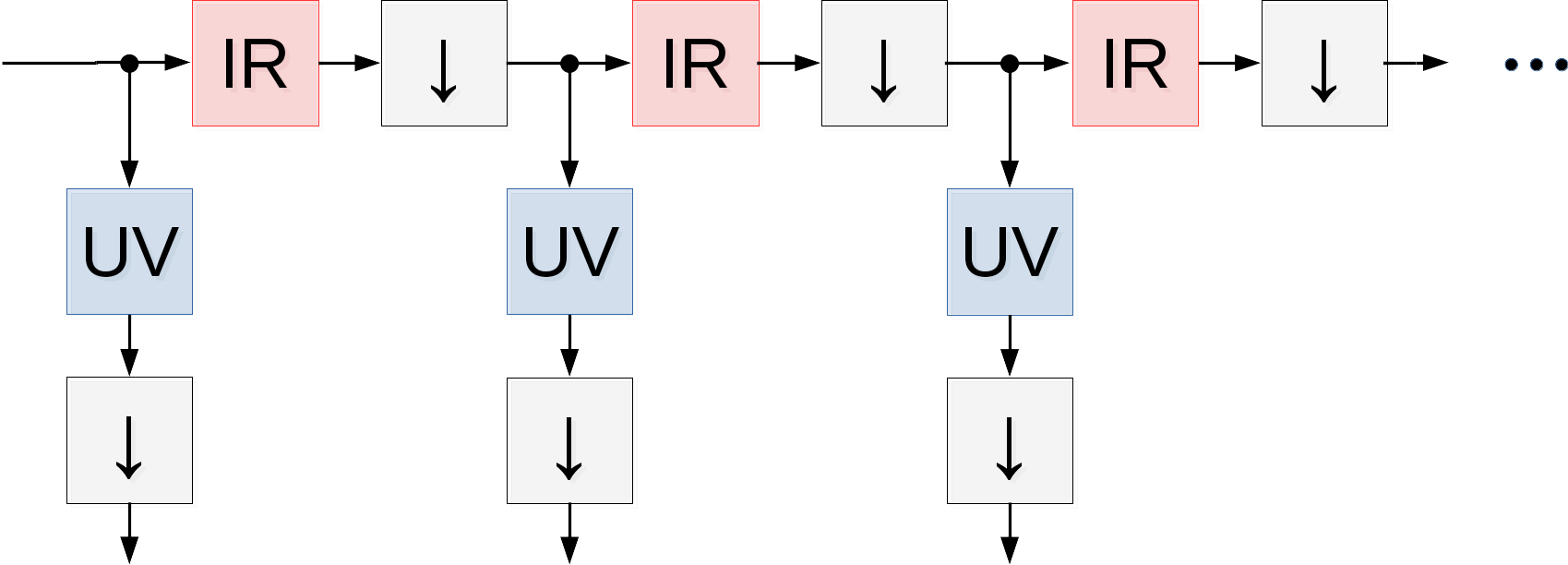}
\vspace{-2mm}
\caption{Wavelet transformation, splitting the incoming signal by a cascade of high-pass filters (UV), low-pass filters (IR), and downsamplers ($\downarrow$). See text.}
\label{fig:wavelet}
\end{figure}

Recently Evenbly and White~\cite{evenbly2016entanglement} showed that discrete wavelet transforms can be used for entanglement renormalization, giving the first analytic example of a MERA that approximates the ground state of a critical Ising model. Wavelet transforms can be understood as localized Fourier transforms, providing a compromise between spectral and real-space resolution, and play an important role in signal processing. 

The basic setup of a wavelet transformation is sketched in Fig.~\ref{fig:wavelet}. The incoming data is split into two identical copies which are passed to a high-pass (UV) and a low-pass (IR) filter. The filtered signals are then processed by downsamplers, which essentially pick up every other data point and therefore halve the degrees of freedom. As shown in the figure, this procedure is iterated several times. In wavelet theory the key point is to choose the filters in such a way that the transformation sequence as a whole is unitary.

\begin{figure}
\centering\includegraphics[width=80mm]{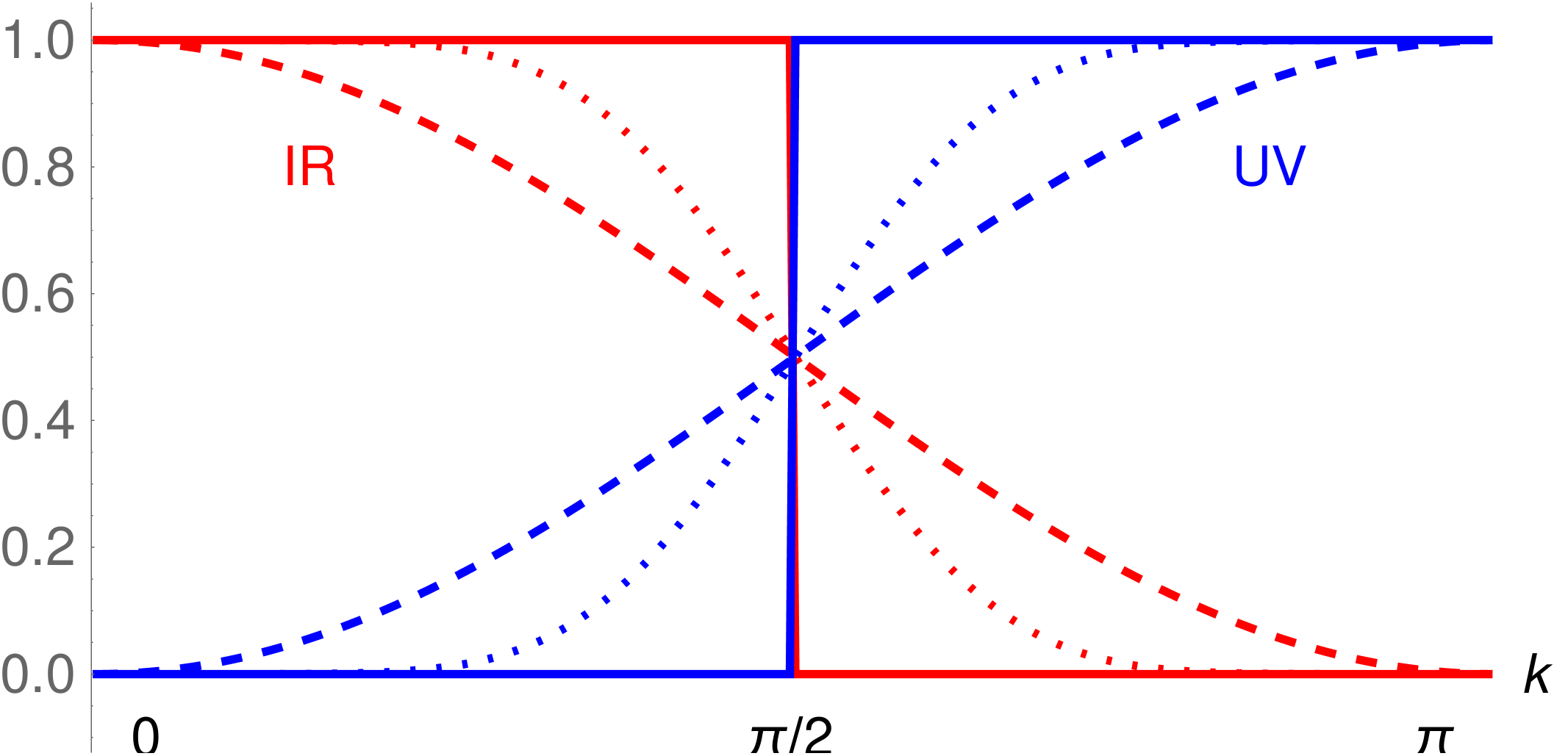}
\vspace{-2mm}
\caption{Filter transfer functions in the range $0 \leq |k| \leq \pi$ for the Daubechies-4 wavelet (dashed line), Daubechies-4 (dotted line), and in the Shannon limit of perfect momentum separation (solid line). }
\label{fig:filter}
\end{figure}

There is a large variety of wavelet transformations with a different trade-off between filter quality and locality. As an example let us consider so-called Daubechies wavelets of order $N$, whose filter functions act on $N$ adjacent lattice sites~\cite{daubechies1992ten}. As demonstrated in Fig.~\ref{fig:filter}, the filter quality of these wavelets increases with $N$ on the expense of locality. In the so-called Shannon limit $N \to \infty$, the wavelet filters separate low and high momenta perfectly (sketched as solid lines in Fig.~\ref{fig:filter}), but then the corresponding filter functions are entirely non-local. 

Although this non-locality may be seen as a disadvantage, the Shannon limit of perfect momentum separation has another significant advantage in the context of renormalization. One observes that for wavelets with a finite range, the overlapping filter functions effectively lead to couplings between the IR and the UV sector, but in the Shannon limit it turns out that both sectors decouple entirely, which is a prerequisite for a practical renormalization scheme. 

If we decided to use finite-range wavelets, we could of course try to remove the remaining couplings between the IR and the UV sectors by additional downstream disentanglers, but it turns out that the resulting compound (wavelet plus disentangler) always reproduces the Shannon limit independent of the initial choice of the wavelet (see Appendix A for further details). Thus we can restrict our analysis to the Shannon limit from the beginning on. 

\begin{figure}
\centering\includegraphics[width=70mm]{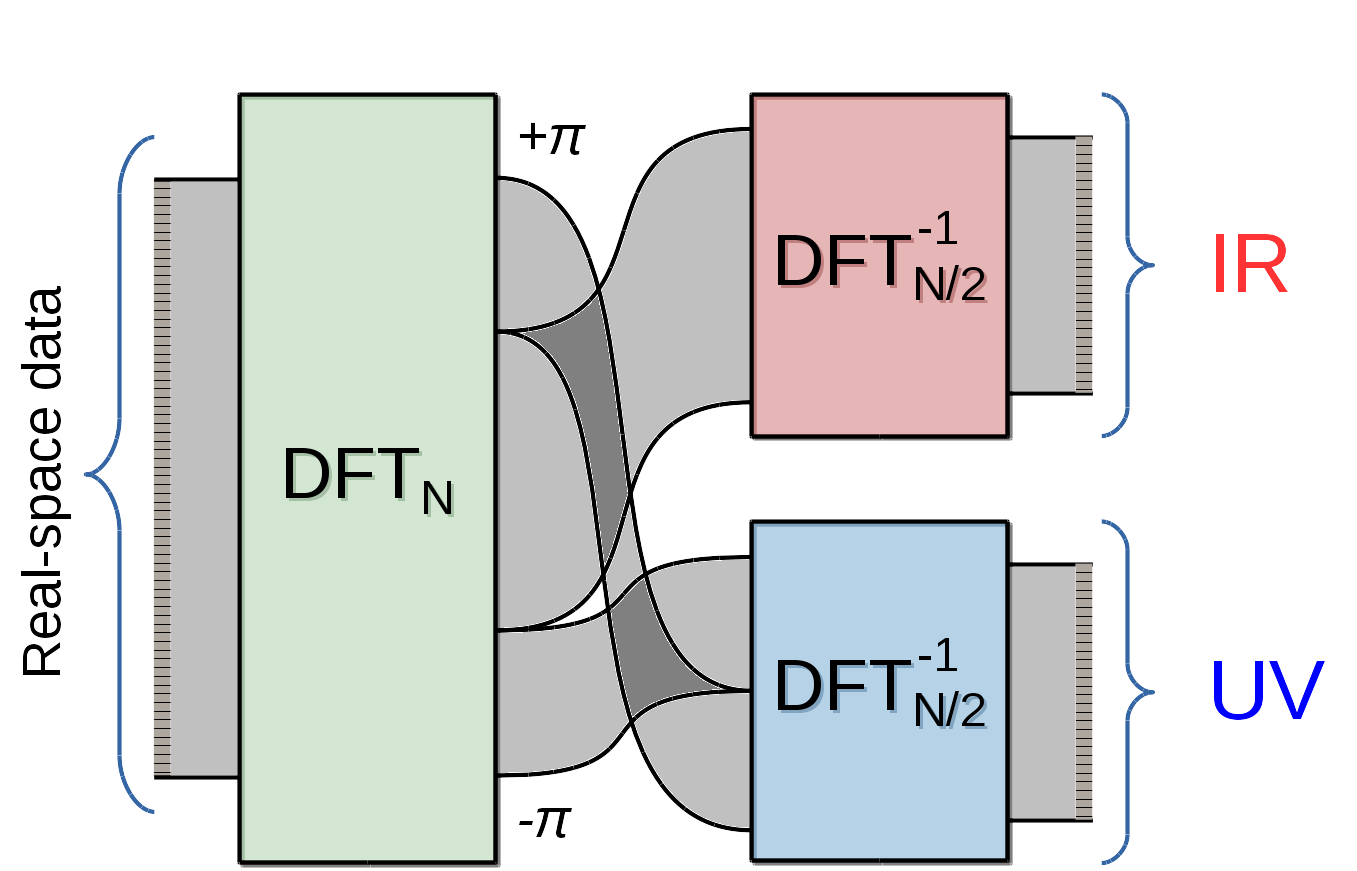}
\vspace{-2mm}
\caption{Perfect momentum separation of a data set. The data is first transformed to momentum space by means of a discrete Fourier transform DFT$_N$. Then the momenta are separated into two halves of low and high momenta, as indicated by the connections bands. These halves are then transformed back to real space by two inverse DFT$_{N/2}$. }
\label{fig:momentum-separation}
\end{figure}

The Shannon limit of perfect momentum separation can be implemented without explicit use of wavelets by a simple sequence of Fourier transformations, as sketched schematically in Fig.~\ref{fig:momentum-separation}. Suppose that we start with a signal or a data set consisting of $N$ discrete values. At first this data set is transformed to $k$-space by a simple discrete Fourier transformed (DFT) of size $N$. Then the output  is divided  'by hand' into two equal halves of low and high momenta, and finally each of the parts is transformed back to real space by an inverse DFT of size $N/2$. This implements a perfect Shannon filter without the explicit use of wavelets and downsamplers. 

\begin{figure}
\centering\includegraphics[width=160mm]{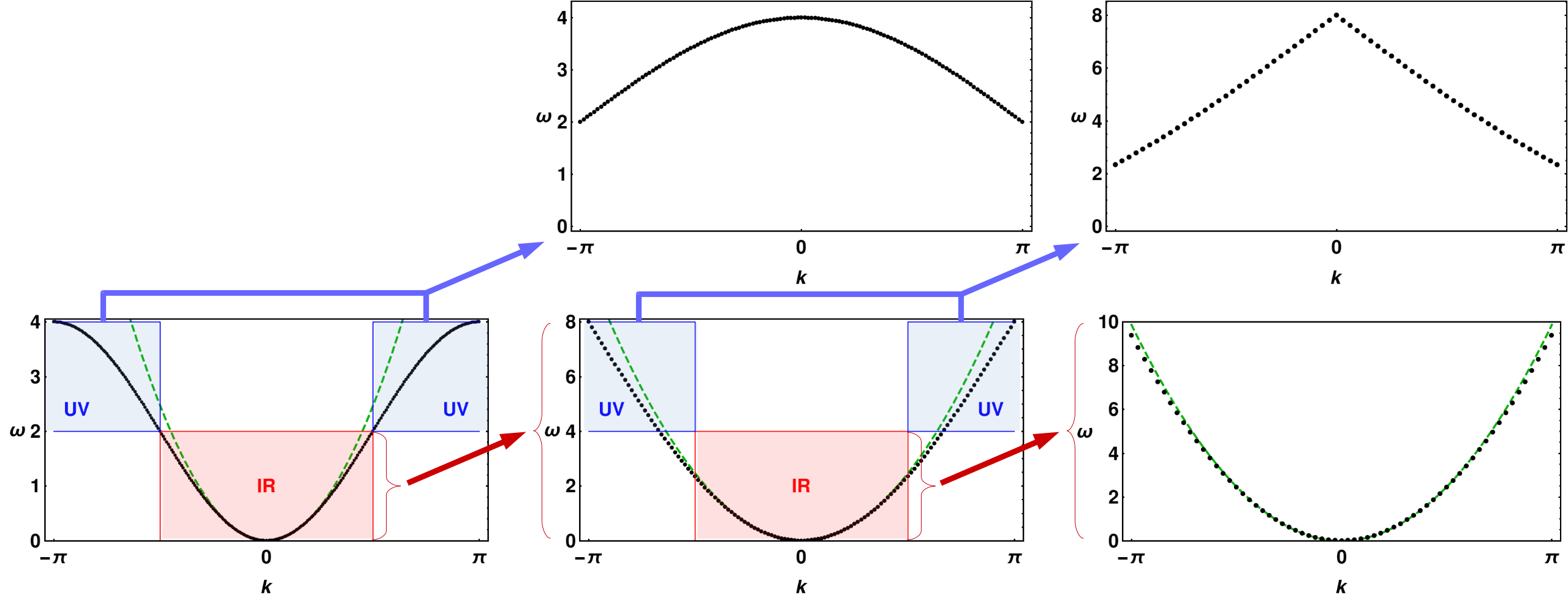}
\caption{Schematic illustration of the renormalization procedure investigated in this paper. The momentum range is divided in to two equal halves, containing the low momenta (IR) and the high momenta (UV). This separation cuts the dispersion relation (solid line) into two pieces. The RG step is iteratively applied to the IR part, magnifying the dispersion relation around $k=0$. This eventually leads to a fixed point where the dispersion relation is exactly quadratic.}
\label{fig:RG-scheme}
\end{figure}

As we are going to show below, the effect of perfect momentum separation applied to free bosonic or fermionic models amounts to cutting the dispersion relation of the excitation modes into two pieces. If we apply this procedure repeatedly to the IR part we obtain an RG scheme which iteratively `zooms' into the low-$k$ regime of the dispersion relation (see Fig.~\ref{fig:RG-scheme}). This process of repeated zooming defines the fixed point of the RG scheme. For example, if we start with a dispersion relation which is parabolic around $k=0$, and if we apply perfect momentum separation several times, the dispersion in the IR sector looks more and more like a parabola, approaching an RG fixed point with a perfectly quadratic dispersion. However, although the UV and IR parts decouple in each step, the couplings \textit{within} each of the parts become increasingly long-ranged under the renormalization flow, and it is intuitively clear that the RG fixed point of a perfectly quadratic dispersion is related to the modified Laplacian~(\ref{longrangelaplacian}) introduced in the previous Section. For the models studied in this paper, this allows us to construct a corresponding clean' variant with long-range interactions which nevertheless exhibits the same universal properties.

\section{Momentum-separating filter functions}
\label{MomentSeparating}

\subsection*{Infinite lattice:}
Let us consider a data set in the form of an infinite-dimensional square-normalizable complex vector $f=\{f_n \in\mathbbm C|n\in\mathbbm Z\}$. This data set can be encoded as a Fourier series
\begin{equation}
\tilde f(k) = \sum_{n\in \mathbbm Z} e^{i k n} f_n
\end{equation}
with a continuum of momenta in the range $k \in [-\pi,\pi]$. This interval is now divided into two equally-sized halves $|k|\leq \frac \pi 2$ (IR) and $\frac \pi 2 < |k| \leq \pi $ (UV), splitting the above Fourier series into
\begin{equation}
  {\tilde f}^{\mathrm{ IR}}(k) \;:=\; \frac{\tilde f\bigl(\tfrac k 2\bigr)} {\sqrt 2} , \qquad
  \tilde f^\mathrm{ UV}(k) \;:=\; \frac{\tilde f\bigl(\tfrac k 2 - \mathrm{sgn}(k) \pi\bigr)}{\sqrt 2}\,,
\end{equation}
where the prefactor of $1/\sqrt{2}$ is included to ensure unitarity. After separating and rescaling the momenta, we apply an inverse Fourier transform to these functions (cf. Fig.~\ref{fig:momentum-separation}), obtaining the real-space vectors
\begin{equation}
    f_m^{\mathrm{IR/UV}} 
    = \frac 1{2\pi}\int_{-\pi}^{+\pi} \mathrm{d} k\, e^{- i k m } \tilde f^{\,\mathrm{IR/UV}}(k) 
    = \sum_{n\in \mathbbm Z} S^{\mathrm{IR/UV}}_{2m-n} f_n,
\end{equation}
where 
\begin{equation}
  S^{\mathrm{IR/UV}}_j 
  = \frac{1}{\sqrt 2 \, \pi} \int_{-\pi}^\pi \mathrm{d} k\, e^{-i k j} \theta\big(\pm (\pi/2 - |k|)\big)
\end{equation}
are the perfect low- and highpass filters. Evaluating these integrals yields
\begin{equation}
\label{eq:Sinf}
S^{\mathrm{IR/UV}}_{j} = \begin{cases}
\frac{\delta_r}{\sqrt{2}} & \mathrm{if} \ j \ \mathrm{even} \\
\pm \frac{\sqrt 2 }{\pi j} \ (-1)^{(j-1)/2}& \mathrm{if} \ j \ \mathrm{odd}
\end{cases}
\end{equation}
or in table form:\\[-4mm]
\begin{center}
  \small
  \begin{tabular}{|c|cccccc|c|cccccc|} \hline
    $j$ & $\cdots$ &  -5 & -4 & -3 & -2 & -1 & \textbf 0 & 1 & 2 & 3 & 4 & 5 & $\cdots$ \\ \hline
    $S^{\mathrm{IR}}_j$ & $\cdots$ & $+\frac{\sqrt{2}}{5\pi}$ & 0 & $-\frac{\sqrt{2}}{3\pi}$ &  0 & $+\frac{\sqrt{2}}{\pi}$ & $\mathbf{\frac{1}{\sqrt{2}}}$ & $+\frac{\sqrt{2}}{\pi}$ & 0 & $-\frac{\sqrt{2}}{3\pi}$ &  0 & $+\frac{\sqrt{2}}{5\pi}$ & $\cdots$  \\ 
    $S^{\mathrm{UV}}_j$ & $\cdots$ &  $-\frac{\sqrt{2}}{5\pi}$ & 0 & $+\frac{\sqrt{2}}{3\pi}$ &  0 & $-\frac{\sqrt{2}}{\pi}$ & $\mathbf{\frac{1}{\sqrt{2}}}$ & $-\frac{\sqrt{2}}{\pi}$ & 0 & $+\frac{\sqrt{2}}{3\pi}$ &  0 & $-\frac{\sqrt{2}}{5\pi}$ & $\cdots$  \\\hline
  \end{tabular}
\end{center}
This transformation can be shown to be unitary in the sense that the block matrix
\begin{equation}
S := (S^{\mathrm{IR}} \oplus S^{\mathrm{UV}})_{n,m} = \binom{S^{\mathrm{IR}}_{2m-n}}{S^{\mathrm{UV}}_{2m-n}}
\end{equation}
satisfies
\begin{equation}
\sum_{l\in\mathbb Z}S^\alpha_{m,l}S^\beta_{n,l} = \delta_{\alpha,\beta}\delta_{m,n}\,. 
\qquad \alpha,\beta \in\{\mathrm{IR,UV}\}
\end{equation}

\subsection*{Finite lattice:}
In analogy to the above, we can define a moment-separating transformation on a square-normalizable \emph{finite} complex vector $f=\{f_0,\ldots,f_{N-1}\} \in \mathbbm C^N $ with $N$ components (for simplicity we assume that $N$ is a power of $2$ and that $N\geq 4$). Here we start with a discrete Fourier transform (DFT)
\begin{equation}
\tilde f_q := \frac{1}{\sqrt{N}} \sum_{n=0}^{N-1} e^{\frac{2\pi \mathrm i n q}{N} } f_n\,,\qquad q\in\{-\tfrac N 2 ,\ldots, \tfrac N 2 -1\}
\end{equation}
where $q$ is the index of the wave numbers $k_q = \frac{2\pi q}{N}$. Again we would like to divide the range of $q$ into two equal halves, separating low and high momenta. However, the two modes $q=\pm \frac N 4$ are exactly on the edge between IR and UV, so it is not clear to which part they belong. We could of course make a choice and assign one of them to the IR part and the other one to the UV part, but then the resulting transformations in real space, $S^{\mathrm{IR}}$ and $S^{\mathrm{UV}}$, would be complex-valued. In order to obtain real transformation matrices, it turns out that we have to assign the symmetric combination of the two modes to the IR part and the antisymmetric contribution to the UV part, i.e., we define
\begin{equation}
\tilde f_{-N/4}^{\,\mathrm{IR}} = \frac 1{\sqrt 2} (\tilde f_{N/4} + \tilde f_{-N/4}), \qquad
\tilde f_{-N/4}^{\,\mathrm{UV}} = \frac 1{\sqrt 2 \ \mathrm i} (\tilde f_{N/4} - \tilde f_{-N/4}),
\end{equation}
and
\begin{equation}
\tilde f_q^{\,\mathrm{IR}} = \tilde f_q, \qquad
\tilde f_q^{\,\mathrm{UV}} = \tilde f_{q-\mathrm{sgn}(q)N/2} , \qquad
q \in \lbrace -\tfrac N 4 + 1, \ldots, \tfrac N 4 - 1 \rbrace.
\end{equation}
Finally, each of these halves is transformed back to real space, giving
\begin{equation}
f_m^{\mathrm{IR/UV}} = \frac 1{\sqrt{N/2}}\sum_{q=-N/4}^{N/4-1} e^{-\frac{4\pi \mathrm i m q}{N}}
\tilde f_q^{\,\mathrm{IR/UV}} = \sum_{n=0}^{N-1}  S^{\mathrm{IR/UV}}_{2m-n} f_n
\end{equation}
with the filters
\begin{equation}
\begin{split}
  S^{\mathrm{IR}}_j 
  &= \frac{2}{N}\cos \frac{\pi j}{2} +
    \frac{\sqrt 2}N \sum_{q=-N/4+1}^{N/4-1} e^{-\frac{2\pi \mathrm i j q}{N}} . \\
  S^{\mathrm{UV}}_j 
  &= \frac{2}{N}\sin \frac{\pi j}{2} +
    \frac{\sqrt 2}N \left(\sum_{q=-N/2}^{-N/4-1} e^{-\frac{2\pi \mathrm i j q}{N}} +\sum_{q=N/4+1}^{N/2-1} e^{-\frac{2\pi \mathrm i j q}{N}}\right) .
\end{split}
\end{equation}
Again, the sums can be evaluated, giving
\begin{equation}
S^{\mathrm{IR/UV}}_{j} =
\begin{cases}
\frac{\delta_j}{\sqrt 2} + \frac{1 \pm (1-\sqrt 2)}N & \mathrm{if} \ j \ \mathrm{even} \\[2mm]
\frac{\sqrt 2 \ (-1)^{(j-1)/2} }N \cot \frac{\pi j}N  & \mathrm{if} \ j \ \mathrm{odd}
\end{cases}
\label{eq:Sfin}
\end{equation}
and the block matrix $S^{\mathrm{IR}} \oplus S^{\mathrm{UV}}$ is unitary and real-valued. As expected, Eq.~(\ref{eq:Sfin}) reduces to Eq.~(\ref{eq:Sinf}) in the infinite lattice limit $N \to \infty$.

Note that the symmetric and antisymmetric combination of the edge modes may lead to unwanted couplings between the IR and the UV part, which have to be eliminated by appropriate transformations. We will come back to this point when discussing the Ising model and the $XY$ chain in Sect.~\ref{IsingSection}.

\section{Renormalization of free bosons}
\label{Free bosons}

\subsection*{Coarse-grained Hamiltonian:}
Let us consider uncharged free bosons of mass $m$ in one space dimension defined by the Hamiltonian
\begin{equation}
\label{eq:bosonicHamiltonian}
  H_B = \frac12 \int \mathrm{d}x \Big[ \Pi^2(x) + m^2 \Phi^2(x)
  + \Big(\frac{\mathrm{d} \Phi(x)}{\mathrm{d}x} \Big)^{\!2} \Big],
\end{equation}
where $\Phi(x)$ is a Hermitean scalar field and $\Pi(x)$ its conjugate, obeying the canonical
commutation relations
\begin{equation}
  [\Phi(x),\Phi(x')]=[\Pi(x),\Pi(x')]=0, \qquad [\Phi(x),\Pi(x')] = \mathrm i \delta(x-x').
\end{equation}
As a starting point for renormalization, we first have to UV-regularize the theory. Introducing the smeared field operators
\begin{equation}
  \Phi(f) = \int_{-\infty}^\infty \mathrm{d}x \, f(x) \Phi(x),\qquad
  \Pi(f) = \int_{-\infty}^\infty \mathrm{d}x \, f(x) \Pi(x)
\end{equation}
with the commutation relations
\begin{align}
  [\Phi(f),\Phi(g)] &= [\Pi(f),\Pi(g)]=0, \\
  [\Phi(f),\Pi(g)] &= \mathrm i\int_{-\infty}^\infty\mathrm{d}x\, f(x) g(x) = i\langle f|g\rangle
\end{align}
and defining an orthonormal set of equidistant real-valued test functions on $L^2(\mathbb{R})$ with $f_n(x) = f_0(x - n)$ and $\langle f_n | f_m \rangle = \delta_{mn}$ one obtains discrete
Hermitean field operators
\begin{equation}
\Phi_n:=\Phi(f_n)\,,\qquad \Pi_n:=\Pi(f_n)
\end{equation}
living on an infinite one-dimensional lattice with lattice spacing $a$. Orthonormality of the test functions implies that
\begin{equation}
  \Phi(x) \approx \sum_{n \in \mathbb Z} \Phi_n f_n(x), \qquad
  \Pi(x) \approx \sum_{n \in \mathbb Z} \Pi_n f_n(x)
\end{equation}
in the distributional sense, which can be used to coarse grain products of the field operators such as
\begin{equation}
  \int \mathrm{d}x\, \Pi^2(x)
  \approx \sum_{n,n' \in \mathbb Z} \Pi_n \Pi_{n'} \int \mathrm{d}x\, f_n(x) f_{n'}(x)
  = \sum_{n \in \mathbb Z} \Pi_n^2.
\end{equation}
If we choose simple rectangular test functions of the form $f_0(x)=\theta(x + 1/2)-\theta(x - 1/2)$, the scheme described above fails for terms involving derivatives of the fields because of discontinuity. However, as shown in the appendix, one can use a renormalization group argument to obtain a coarse-grained version of the Hamiltonian~(\ref{eq:bosonicHamiltonian}), reading
\begin{equation}\label{eq:H1}
  H = \frac 12 \sum_{n \in \mathbb Z} \Big[ \Pi_n^2 +   m^2 \Phi_n^2 \Big]
  - \frac 12 \sum_{n,n'\in \mathbb Z} \Phi_n \Delta_{n-n'} \Phi_{n'},
\end{equation}
where $\Delta$ denotes the standard discretized Laplacian in one dimension (cf. Eq.~(\ref{discreteSchroedinger})):
\begin{equation}\label{discreteLaplacian}
  \Delta_j=\begin{cases} 
    -2 & \text{if } j=0 \\[-2mm]
    +1 &\text{if } j=\pm 1 \\[-2mm] 
    0 & \text{otherwise} \end{cases}
\end{equation}
Similarly, we can define the bosonic Hamiltonian on a \textit{finite} ring of $N$ sites by
\begin{equation}\label{eq:H2}
  H = \frac 12 \sum_{n=0}^{N-1} \Big[ \Pi_n^2 +   m^2 \Phi_n^2 \Big]
  - \frac 12 \sum_{n,n'=0}^{N-1} \Phi_n \Delta_{n-n'} \Phi_{n'},
\end{equation}
assuming periodic boundary conditions for $\Delta_j$. 

\subsection*{Renormalization scheme:}

Denoting by $\Phi=\{\Phi_n\}$ and $\Pi=\{\Pi_n\}$ the vectors of all field operators on the chain, the coarse-grained Hamiltonian that we use as the starting point for renormalization can be written in a compact form by
\begin{equation}
\label{initialBosonicHamiltonian}
H \;=\; \frac12\begin{bmatrix} \Pi \\ \phi \end{bmatrix}^\dagger 
\begin{bmatrix} \mathbbm 1 & \\ & m^2 \mathbbm 1 - \Delta \end{bmatrix}
\begin{bmatrix} \Pi \\ \phi \end{bmatrix}
\end{equation}
with the short-range Laplacian $\Delta$ defined in Eq.~(\ref{discreteLaplacian}). At this point, we apply the moment-separating filters $S=S^{\mathrm{IR}} \oplus S^{\mathrm{UV}}$ introduced in Sect.~\ref{MomentSeparating}, and define the renormalized field operators via
\begin{equation}
\begin{bmatrix} \Pi_\text{IR} \\  \Pi_\text{UV} \end{bmatrix} = S \Pi\,,\qquad 
\begin{bmatrix} \Phi_\text{IR} \\  \Phi_\text{UV} \end{bmatrix} = S \Phi \,.
\end{equation}
Since the transformation matrix $S$ is unitary, it is clear that these renormalized operators also obey bosonic commutation relations. With this definition we can rewrite $H$ as
\begin{equation}
H=
\begin{bmatrix} \Pi_\text{IR} \\  \Pi_\text{UV} \\ \Phi_\text{IR} \\  \Phi_\text{UV} \end{bmatrix} ^\dagger
\underbrace{\begin{bmatrix} S & \\ & S \end{bmatrix} 
\begin{bmatrix} \mathbbm 1 & \\ & m^2 \mathbbm 1 - \Delta \end{bmatrix}
\begin{bmatrix} S & \\ & S \end{bmatrix} ^\dagger}_{{\scriptscriptstyle =\begin{bmatrix} \mathbbm 1 & \\ & m^2 \mathbbm 1 - S \Delta S^\dagger \end{bmatrix}}}
\begin{bmatrix} \Pi_\text{IR} \\  \Pi_\text{UV} \\ \Phi_\text{IR} \\  \Phi_\text{UV} \end{bmatrix} 
\end{equation}
where the transformed Laplacian has the structure
\begin{equation}
S \Delta S^\dagger = 
\begin{bmatrix} 
S^\text{IR} \Delta {S^\text{IR}}^\dagger & 
S^\text{UV} \Delta {S^\text{IR}}^\dagger \\
S^\text{IR} \Delta {S^\text{UV}}^\dagger & 
S^\text{UV} \Delta {S^\text{UV}}^\dagger
\end{bmatrix}\,.
\end{equation}
A key point of this work is the observation that for perfect momentum separation this matrix has a block-diagonal form, i.e.,
\begin{equation}
S^\text{UV} \Delta {S^\text{IR}}^\dagger = S^\text{IR} \Delta {S^\text{UV}}^\dagger = 0.
\end{equation}
This can be verified most easily in momentum space, where the translation-invariant operator $\Delta$ is diagonal, and realizing that the two filters have no common overlap. 

The block-diagonal form implies that the Hamiltonian \textit{decouples} into a direct sum 
\begin{equation}
  H=H_\text{IR}\oplus H_\text{UV}
\end{equation}
with
\begin{equation}
H_\text{IR}=\begin{bmatrix} \Pi_\text{IR} \\  \Phi_\text{IR}  \end{bmatrix} ^\dagger
\begin{bmatrix} \mathbbm 1 & \\ & m^2 \mathbbm 1 - S^\text{IR}\Delta{S^\text{IR}}^\dagger \end{bmatrix}
\begin{bmatrix} \Pi_\text{IR} \\ \Phi_\text{IR} \end{bmatrix} 
\end{equation}
and an analogous expression for $H_\text{UV}$. Note that this decoupling relies on perfect momentum separation as well as the preservation of the commutation relations. 

We now discard the ultraviolet part and consider the infrared part $H^{(1)}:=H_\text{IR}$ as the renormalized Hamiltonian. Repeating this procedure many times we obtain a sequence of Hamil\-tonians
\begin{equation}
H=H^{(0)}\to H^{(1)}\to H^{(2)} \to \ldots \to H^{(r)}\to \ldots
\end{equation}
Each renormalization step divides the dimension of the Hamiltonian by $2$. Therefore, if we start with a finite lattice, it is useful to choose $N$ as a power of $2$. For example, if we start with $N=2^\ell$ lattice sites, the Hamiltonian $H^{(r)}$ after $r$ renormalization steps is a $2^{l-r}\times 2^{l-r}$ matrix, while on infinite lattices $H^{(r)}$ is an infinite-dimensional square matrix on all levels.

\subsection*{Exact renormalization of the discretized Laplacian:}
In order to analyze the RG flow, we have to understand how the discrete Laplacian $\Delta$ changes under repeated action of $S$. To this end we recursively define the $r$-fold renormalized Laplacian as
\begin{equation}
\Delta^{(r)} := S^\text{IR}\Delta^{(r-1)}{S^\text{IR}}^\dagger\,,\qquad \Delta^{(0)}:=\Delta\,,
\end{equation}
where the size of the transformation matrix $S_\text{IR}$ is different in each step. A straight-forward calculation based on Fourier techniques (not shown here) gives the exact result for the $r$-fold renormalized Laplacian acting on a lattice of $N$ sites:
\begin{equation}
\label{Deltar}
  \Delta^{(r)}_{j} =  
  \begin{cases}
    \frac{\frac{2}{N} \sin \left( 2^{-r} \pi \right)}{\tan \left(  2^{-r}\pi/N\right)}-2
    & \text{if } j=0    \\[2mm]
    \frac{\frac{2}{N} (-1)^{j} \sin \left(  2^{1-r}\pi/N\right) \sin \left(  2^{-r} \pi\right)}
    {\cos \left( 2 j \pi/N\right)-\cos\left( 2^{1-r} \pi/N \right)}
    &\text{if } j\neq 0  
\end{cases}
\end{equation}
On an infinite lattice this expression reduces to
\begin{equation}
  \lim_{N\to \infty} \Delta^{(r)}_{j} =  
  \begin{cases}
    \frac{2^{1+r} \sin \left( 2^{-r} \pi \right)}{\pi} -2
    & \text{if } j=0    \\[2mm]
    \frac{2^{1+r} (-1)^{j}  \sin \left(  2^{-r} \pi\right)}
    {\pi(1-4^r j^2)}
    & \text{if } j\neq 0  
\end{cases}
\end{equation}
Analyzing the solution (\ref{Deltar}) for large values of $r$ one finds that in each renormalization step, which essentially doubles the length scale, the Laplacian picks up the factor $1/4$. Thus, if we rescale the Laplacian by $4^r$ and take $r \to \infty$ we obtain the IR fixed point
\begin{equation}
\label{FixedPointLaplacian}
  \Delta^*_j = \lim_{r \to \infty} 4^r \Delta^{(r)}_{j} =  
  \begin{cases}
    -\frac{\pi^2}{3} -\frac{2\pi^2}{3 N^2} & \text{if } j=0  \\[2mm]
    \frac{2 \pi^2 \, (-1)^{1+j}}
    {N^2 \sin^2\left(  \pi j/N\right)}  & \text{if } j\neq 0  
  \end{cases}
\end{equation}
which in the limit $N \to \infty$ reduces to the Laplacian $\Delta^*_{j}$ on the infinite lattice defined in Eq.~(\ref{longrangelaplacian}). The expression given in (\ref{FixedPointLaplacian}) is the optimal' Laplacian on a finite lattice with $N$ sites which exhibits a clean quadratic dispersion. It can be understood as the ``correct'' discretization of the Laplacian, corresponding to our choice of RG scheme. Indeed, as shown in \ref{AA}, different choices of wavelet filters correspond to different fixed points, which can be directly found from the filter coefficients. 

As for the renormalization of the Hamiltonian, a non-trivial RG fixed point can only be obtained if we rescale the Laplacian by a factor $4^r$. Therefore, the free boson Hamiltonian on a lattice with $N$ sites obtained by $r$-fold renormalization of the original Hamiltonian on $2^rL$ sites, is given by
\begin{equation}
H^{(r)}=\begin{bmatrix} \Pi \\  \Phi  \end{bmatrix} ^\dagger
\begin{bmatrix} \mathbbm 1 & \\ & m^2 \mathbbm 1 - 4^r \Delta^{(r)} \end{bmatrix}
\begin{bmatrix} \Pi \\ \Phi \end{bmatrix} \,.
\end{equation}
This Hamiltonian can be diagonalized via Fourier transformation, giving
\begin{equation}
H^{(r)}=\sum_{q=-N/2}^{N/2-1} \omega_q^{(r)} \Bigl(\eta_q^\dagger\eta_q + \tfrac12 \Bigr)
\end{equation}
with a bosonic number operators $\eta_q^\dagger\eta_q$ and the excitation energies
\begin{equation}
\label{wr}
 \omega_q^{(r)} = \sqrt{m^2 + 4^r\bigl[2-2\cos\bigl( 2^{-r} k_q\bigr)\bigr]}\,,
\end{equation}
where $k_q = 2\pi q/N $. In the IR limit, this expression converges to
\begin{equation}
\omega_q^* = \lim_{r \to \infty} \omega_q^{(r)} = \sqrt{m^2 + k_q^2}\,.
\end{equation}
which coincides \emph{exactly} with the continuum solution.

\subsection*{Example: Heat capacity}
In order to demonstrate that the long-range Hamiltonian at the RG fixed point approximates the continuum limit more accurately that the unrenormalized short-range Hamiltonian we compute the heat capacity
%
\begin{figure}
\centering\includegraphics[width=150mm]{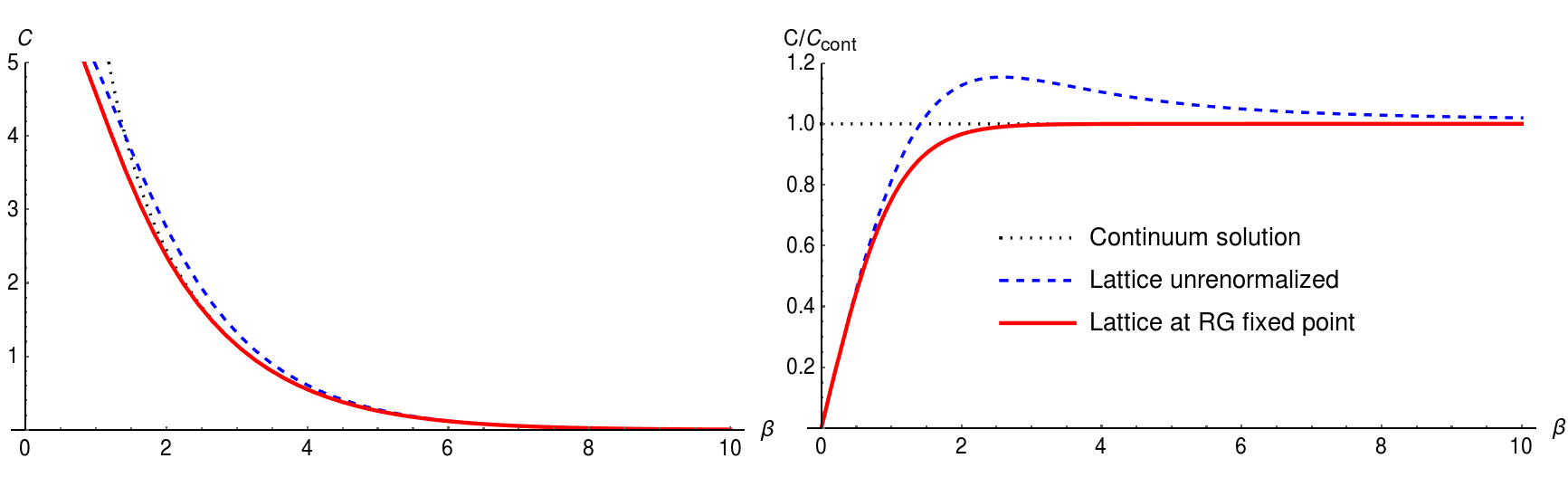}
\label{fig:specific-heat}
\caption{Left: Specific heat of free bosons in one dimension with the mass $m=1$ as a function of the inverse temperature $\beta=1/T$, calculated for the continuum model, the discretized short-range model, and the long-range model at the RG fixed point. Right: For better visualization of the deviations, we show the same data divided by the continuum solution. }
\end{figure}
%
\begin{equation}
C=\beta^2 \, \frac{\partial^2}{\partial \beta^2} \,\ln Z_\beta
\end{equation}
where 
\begin{equation}
\ln Z_\beta = \ln \sum_{\{n_q\}} e^{-\beta H} = \sum_q \ln \sum_{n=0}^\infty e^{-\beta \omega_q (n+1/2)} =
\sum_q \ln\frac{1}{e^{\beta\omega_q/2}-e^{-\beta\omega_q/2}} .
\end{equation}
giving
\begin{equation}
C= \beta^2\sum_q \frac{\omega_q^2 }{(e^{\beta\omega_q/2}-e^{-\beta\omega_q/2})^2}\,.
\end{equation}
In the original continuum model, the specific heat is given by
\begin{equation}
C_\text{cont} =  \beta^2 \int_{-\infty}^{+\infty} \textrm d k \, \frac{m^2+k^2}{\Bigl(e^{\frac12\beta\sqrt{m^2+k^2}}-e^{-\frac12\beta\sqrt{m^2+k^2}}\Bigr)^2}
\end{equation}
On a lattice with short-range Laplacian, the corresponding non-renormalized expression in the thermodynamic limit $L\to\infty$ reads
\begin{equation}
C_\text{lattice} =  \beta^2 \int_{-\pi}^{+\pi} \textrm d k \, \frac{m^2+2-2\cos k}{\Bigl(e^{\frac12\beta\sqrt{m^2+k^2}}-e^{-\frac12\beta\sqrt{m^2+k^2}}\Bigr)^2}\,.
\end{equation}
At the renormalization fixed point, the deformed dispersion relation in the denominator turns into a quadratic one:
\begin{equation}
C^*_\text{lattice} =  \beta^2 \int_{-\pi}^{+\pi} \textrm d k \, \frac{m^2+k^2}{\Bigl(e^{\frac12\beta\sqrt{m^2+k^2}}-e^{-\frac12\beta\sqrt{m^2+k^2}}\Bigr)^2}
\end{equation}
In Fig.~\ref{fig:specific-heat} these expressions are plotted as functions of the inverse temperature $\beta$. As expected, both lattice approximations, the unrenormalized and the renormalized one, differ from the continuum model for high temperatures, where high momenta beyond the Brillouin zone become relevant. It is also clear that all curves converge in the limit of low temperatures, where the excitation modes are much longer than the lattice spacing. In the intermediate range, however, the figure demonstrates that the long-range Hamiltonian at the RG fixed point approximates the continuum model much better than the short-range model from where we started. 
\subsection*{Example: Massless two-point correlation function}
As a second example, we study the two-point equal time correlator $G(x) = \langle \Omega | \Phi(n) \Phi(n+x) | \Omega \rangle$ for $m=0$. To this end, let us begin with the momentum space continuum Hamiltonian 
\[
   H = \frac 12 \int \mathrm{d}k \, \Big[ \Pi(-k)\Pi(k) + k^2 \Phi(-k)\Phi(k) \Big].
\]
In the usual canon, we define annihilation operators
\[
  a_k \equiv \sqrt{\frac{|k|}2} \Big[\Phi(k) + i |k|^{-1} \Pi(k) \Big],
\]
annihilating the vacuum $a_k |\Omega\rangle = 0$. This allows to rewrite the Hamiltonian as
\begin{equation}\label{eq:diag-cont-hamiltonian}
  H = \int \mathrm{d}k \, |k| a_k^\dagger a_k^{} + \textrm{const.}
\end{equation}
and we see that $|\Omega\rangle$ is indeed the ground state. We directly obtain
\begin{equation}\label{eq:cont-kspace-corr}
  \langle \Omega | \Phi(k)\Phi(k') | \Omega \rangle = \frac 1{2\sqrt{|kk'|}} \langle \Omega | a^{}_{k} a^\dagger_{-k'} | \Omega \rangle = \frac {\delta(k+k')}{2 |k|},
\end{equation}
or, in real space,
\begin{equation} \label{eq:cont-realspace-corr}
  G_{\textrm{cont}}(x, \Lambda) = \frac 1{2\pi} \int_{1/\Lambda}^{\infty} \mathrm{d} k \, \frac{\cos k x}k \sim \frac 1{2\pi} [\log \Lambda - \log x - \gamma],
\end{equation}
where $\gamma$ is the Euler-Mascheroni constant and we introduced an IR cutoff $\Lambda$ to extract the overall divergent part of the integral. The UV divergence at $x \to 0$ remains.

\begin{figure}
\centering\includegraphics[width=80mm]{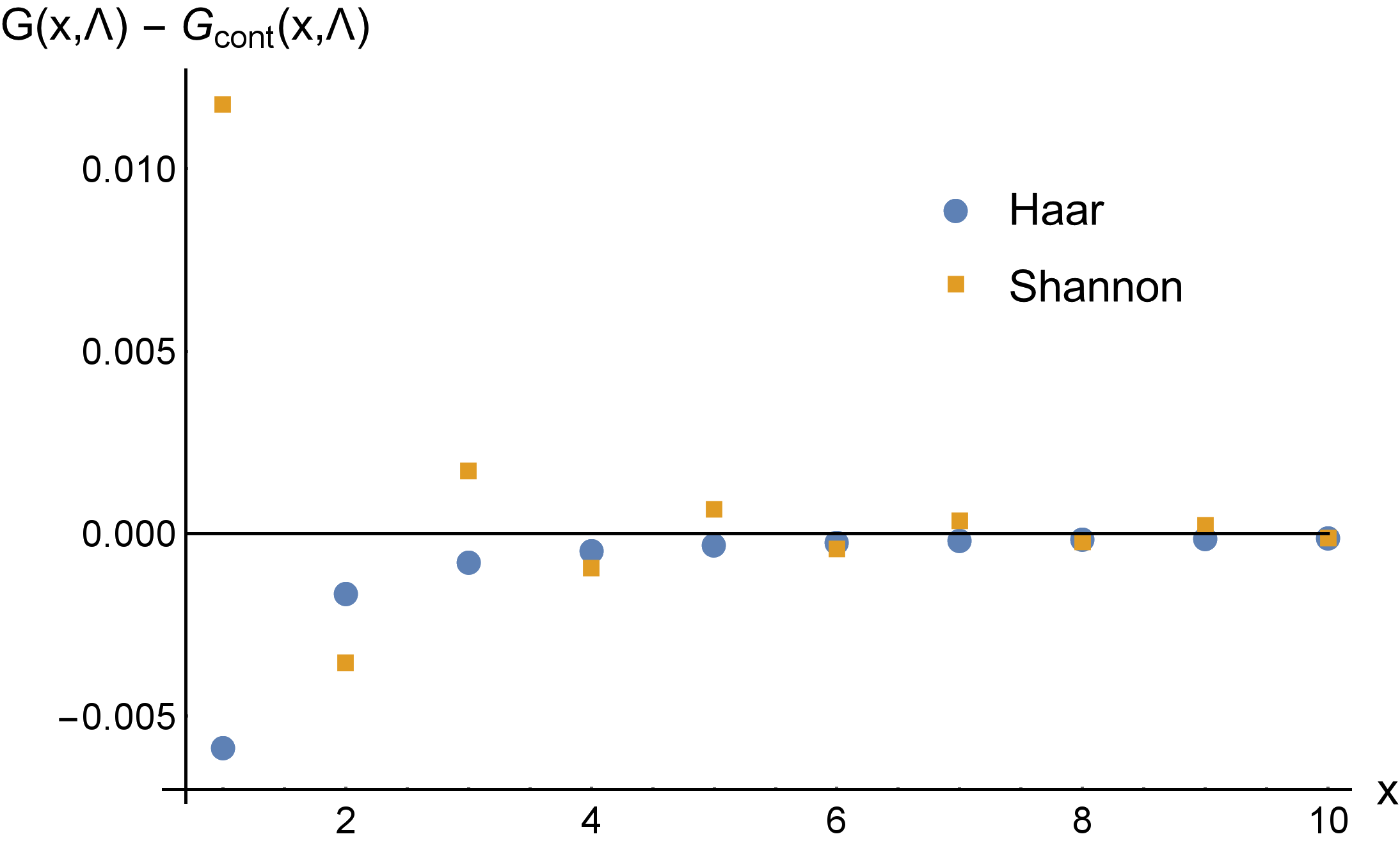}
\label{fig:boson-correlator}
\caption{Deviation of the two-point function $G(x, \Lambda)$ in the discrete theory from its continuum analog, according to Eqs.~(\ref{eq:cont-realspace-corr}),~(\ref{eq:disc-realspace-corr}),~and~(\ref{eq:disc-realspace-corr-fp}). The non local fixed point (yellow) oscillates around the continuum solution at the scale of a single lattice spacing, as is to be expected from the sharp cutoff at $k = \pm \pi$.}
\end{figure}

The above has to be compared with the corresponding two-point correlation functions on the infinite lattice. Here, we can repeat the calculations in truncated momentum space $k \in [-\pi,\pi]$ for the lattice dispersion relations $\Delta^{(r)}(k)$. More specifically, in the short-range case $\Delta(k)=\sqrt{2-2\cos k}$, we get
\begin{equation} \label{eq:disc-realspace-corr}
G_{\textrm{lattice}}(x, \Lambda) = \frac 1{2\pi} \int_{1/\Lambda}^\pi \textrm d k \, \frac{\cos k x}{2 \sin k/2},
\end{equation}
while the same quantity at the RG fixed point $\Delta^*(k) = |k|$ is
\begin{equation} \label{eq:disc-realspace-corr-fp}
  G^*_{\textrm{lattice}}(x, \Lambda) = \frac 1{2\pi} \int_{1/\Lambda}^{\pi} \mathrm{d} k \, \frac{\cos k x}k.
\end{equation}
As we see in Fig.~\ref{fig:boson-correlator}, $G^*_{\textrm{lattice}}$ oscillates around the continuum solution -- we identify this as a ringing artifact, originating in the sharp cutoff at $|k|=\pi$. 
%
%

\section{Renormalization of free fermions}

As a second application, we consider massive Dirac fermions in 1+1 dimensions with the equal-time anticommutation relations
\begin{equation}
\begin{split}
\{\psi^a(t,x),\psi^b(t,y)\}&=\{{\psi^a}^\dagger(t,x),{\psi^b}^\dagger(t,y)\}=0\,,\\ 
\{\psi^a(t,x),{\psi^b}^\dagger(t,y)\} &=\delta^{ab}\delta(x-y)\,.
\end{split}
\end{equation}
The fermions evolve according to the free Lagrange density
\begin{equation}
\mathcal L = \bar \psi ( i \gamma^\mu \partial_\mu -m )\psi\,,
\end{equation}
where $m$ is the mass and $\bar\psi=\psi^\dagger \gamma^0$. Defining the conjugate momenta
\begin{equation}
\Pi(x) \;=
i \psi^\dagger(x)
\end{equation}
with the equal-time anticommutation relations
\begin{equation}
\{\psi^a(t,x),\psi^b(t,y)\}=\{\Pi^a(t,x),\Pi^b(t,y)\}=0\,,\quad \{\psi^a(t,x),\Pi^b(t,y)\}=i \delta^{ab}\delta(x-y)
\end{equation}
the corresponding Hamiltonian density reads
\begin{equation}
\mathcal H \;=\; \Pi \partial_0\psi - \mathcal L \;=\; \Pi\bigl( -\gamma^0\gamma^1\partial_1 - i m \gamma^0)\psi
\end{equation}
In the chiral basis $\gamma^0=\sigma^x, \gamma^1=i\sigma^y$, the Hamiltonian is given by
\begin{align}\label{eq:fermion-hamiltonian}
\mathcal H = \Pi (\sigma^z\partial_1-im\sigma^x)\psi
\end{align}

As in the bosonic case we consider a discrete set of smeared field operators
\begin{equation}
\psi_n = \int_{-\frac12}^{\frac12} \mathrm{d}x\, \psi(x-n)\qquad
\Pi_n = \int_{-\frac12}^{\frac12} \mathrm{d}x\, \Pi(x-n) \, \textrm{d}x
\end{equation}
with the anticommutation relations
\begin{equation}
\lbrace \psi^a_n,\psi^b_{n'}\rbrace = \lbrace \Pi^a_{n},\Pi^b_{n'} \rbrace = 0\,, \,\qquad
\lbrace \psi^a_{n},\Pi^b_{n'} \rbrace = i\delta^{ab}\delta_{nn'}\,.
\end{equation}
In terms of these operators, the bare Hamiltonian, which serves as the starting point of the RG procedure, is given by
\begin{equation}
H \;=\; \sum_{n,n'}\Pi_n \sigma^z \partial_{n-n'} \psi_{n'} \,-\, i m \sum_n \Pi_n \sigma^x \psi_n\,.
\end{equation}
Similarly to for the lattice Laplacian \eqref{discreteLaplacian}, a naive lattice discretization of the first derivative can be given by
\begin{equation}
\label{discreteGradient}
\partial_n=\begin{cases} 
\pm 1/2 &\text{if } n = \pm 1 \\
0 & \text{otherwise.} \end{cases}
\end{equation}

As is well known, its Fourier transform is
\begin{align}\label{grad}
\partial(k)= i \sin(k),
\end{align}
which becomes linear only for $|k| \rightarrow 0$, but otherwise introduces a number of inconvenient lattice effects. 

\subsection*{Renormalization via Shannon wavelet:}

Written in matrix form, the Hamiltonian is
\begin{equation}
\label{initialFermionicHamiltonian}
H \;=\; \frac12\begin{bmatrix} \Pi \\ \psi \end{bmatrix}^T 
\begin{bmatrix}  & -im\sigma^x \mathbbm 1 + \sigma^z \partial \\ &  \end{bmatrix}
\begin{bmatrix} \Pi \\ \psi \end{bmatrix}
\end{equation}
By the same reasoning as in the bosonic section, an RG step that decouples exactly the UV and IR components of the fields at each iteration will consequently decouple the Hamiltonian, the IR contribution being
\begin{equation}
H_\text{IR}=\begin{bmatrix} \Pi_\text{IR} \\  \Psi_\text{IR}  \end{bmatrix} ^T
\begin{bmatrix}  & -im \sigma^x \mathbbm 1 + \sigma^z S^{\text{IR}}\partial S^{{\text{IR}}\, T}  \\ &  \end{bmatrix}
\begin{bmatrix} \Pi_\text{IR} \\ \Psi_\text{IR} \end{bmatrix} 
\end{equation}

Let us now analyze the behavior of the lattice operator $\partial(k)$ in \eqref{grad} under the action of $S^{\text {IR}}$. By construction, after $r$ RG steps, the Fourier transformed derivative operator is the original operator `zoomed' by a factor of $2^r$,
\begin{align}\label{}
\partial^{(r)}(k):= i \sin \left( 2^{-r}k \right)
\end{align}
To get it in real space, we simply Fourier transform back to obtain
\begin{equation}\label{partialr}
\partial^{(r)}_{n} = \frac i{2\pi} \int_{-\pi}^\pi \mathrm{d} k\, e^{-i k n} \sin \left( 2^{-r}k \right) = \frac{(-1)^n}{2^r n - (2^r n)^{-1}} \frac{\sin \left( 2^{-r} \pi \right)}{2^{-r} \pi}.
\end{equation}
As we go to the IR $r\rightarrow \infty$, $\eqref{partialr}$ scales as $2^{-r}$, as it should for a first derivative operator of scaling dimension one. The fixed point is
\[
\partial^*_n=\lim_{r\rightarrow \infty} 2^r \partial^{(r)}_n = \frac{(-1)^n}n. 
\]

Again, we see that the improved lattice Shannon fixed point operator does not have compact support, but falls linearly with distance and is thus highly non-local. The dispersion relation for the renormalized fermion is given by (the absolute value of) the eigenvalues of the matrix 
\[
  -im\sigma^x \mathbbm 1 + \sigma^z \partial^{(r)}(k), 
\]
which evaluates to
\[
E^{(r)}(k) = \sqrt{m^2 +\left| 2^r \partial_k^{(r)} \right|^2} = \sqrt{m^2 + \left(\frac{\sin 2^{-r}k}{2^{-r}}\right)^{\!\!2}}.
\]
In the IR limit $r \to \infty$, this yields the exact continuum dispersion
\begin{align}\label{gradFP}
E^*(k)=\sqrt{m^2+k^2}
\end{align}
in analogy to the bosonic case \eqref{wr}.

\subsection*{On Fermion doubling and non-locality}
One of the most fundamental and ubiquitous results in lattice field theory is the Nielsen-Ninomiya theorem\cite{Nielsen:1981hk,Kaplan:2009yg}. It states that, given the following action for massless fermions in momentum space,
\[
\int_{-\pi/a}^{\pi/a} d^{2d}k\, \bar \psi(-k) D(k) \psi_k
\]
on a lattice of spacing $a$, the operator $D$ cannot satisfy the following four properties simultaneously:
\begin{enumerate}
\item $D(k)$ is a periodic analytic function of $k^\mu$
\item $D(k) \sim \gamma^\mu k_\mu$ for $a|k|\ll 1$
\item $D(k)$ is invertible for $k_\mu\neq 0$
\item $\{ \gamma^5,D(k) \}=0$
\end{enumerate}


In $1+1$ dimensions, chiral symmetry implies that the two components of the Dirac spinor $\psi$ decouple as
\[
H = \frac1{2\pi} \int_{-\pi/a}^{\pi/a} \mathrm{d}k\, D(k) \Big[[\psi^1(k)]^\dagger \psi^1(k) - [\psi^2(k)]^\dagger \psi^2(k) \Big],
\]
where $D(k)$ now is a \emph{scalar} function. This Hamiltonian is easily diagonalized: We define annihilation operators
\[
  c_k =
  \begin{cases}
    \psi^1(k) &\text{for } D(k) > 0 \\
    [\psi^1(k)]^\dagger &\text{for } D(k) < 0
  \end{cases}, \qquad
  b_k =
  \begin{cases}
    [\psi^2(k)]^\dagger &\text{for } D(k) > 0 \\
    \psi^2(k) &\text{for } D(k) < 0
  \end{cases}
\]
so that 
\begin{equation}\label{eq:diagonal-hamiltonian}
  H = \frac1{2\pi} \int_{-\pi/a}^{\pi/a} \mathrm{d}k\, |D(k)| \Big[c^\dagger_k c^{}_k + b^\dagger_k b^{}_k\Big] + \mathrm{const.}
\end{equation}
As in the bosonic case, the vacuum is annihilated by the $c_k$ and $b_k$. The low-energy excitations are created by $c^\dagger_k$ and $b^\dagger_k$, where $D(k) \approx 0$. 

The fermion doubling problem is merely the statement that if $D(k)$ is periodic in the Brillouin zone $[-\pi/a,\pi/a]$, then for every time that $D(k)$ crosses $0$ from below, it must cross it somewhere else from above. This implies that on a lattice, the number of low-energy excitations (i.e., particles) of types $c$ and $b$ must be \emph{even}. In contradistinction, a continuum theory with, e.g., $D(k) = k$ allows for a \emph{single} particle of type $c$ and $b$, respectively. There are many well known methods to eliminate the unwanted doublers on the lattice, by avoiding any combination of the above assumptions, e.g., staggered, Wilson domain wall and overlap fermions \cite{Kogut:1974ag,Chandrasekharan:1998wg,Jansen:1994ym,Creutz:1984mg}. Let us now examine this issue within our approach.

\subsection*{Mean occupation number (and heat capacity)}
As is clear from the above subsections, our construction clearly violates the first assumption of the Nielsen-Ninomiya theorem: the RG fixed point of the lattice operator $\partial^*_{\textrm{lattice}}(k)$ is non-analytic in the Brillouin zone. To verify that it is also free of doubler fermions, we compute the mean occupation number in the canonical ensemble. To this end, consider the partition function
\[
  Z[\beta, \mu_c, \mu_b] = \mathrm{Tr}\, \exp \Big[ -\beta H + \mu_c N_c + \mu_b N_b \Big],
\]
where
\[
  N_c = \frac{1}{2\pi} \int_{-\pi}^\pi \mathrm{d}k\, c^\dagger_k c^{}_k \qquad N_b = \frac{1}{2\pi} \int_{-\pi}^\pi \mathrm{d}k\, b^\dagger_k b^{}_k.
\]
Imposing a finite size $L$ and using the diagonal form of the Hamiltonian~(\ref{eq:diagonal-hamiltonian}), the above trace reduces to
\[
  Z[\beta, \mu_c, \mu_b] = \prod_{e^{i k L} = 1} \mathrm{Tr}\, \big[ e^{(\mu_c - \beta |D(k)|) c^\dagger_k c^{}_k} \big] \mathrm{Tr}\, \big[ e^{(\mu_b - \beta |D(k)|) b^\dagger_k b^{}_k} \big].
\]
We can therefore evaluate the mean occupation number $n_c = N_c/L$ in the thermodynamic limit $L \to \infty$ as
\[
  \langle n_c \rangle_\beta = \left.\frac{\mathrm d}{\mathrm d \mu_c}\right|_{\mu_c,\mu_b=0} \!\!\!\!\!\!\! \frac{\log Z[\beta, \mu_c, \mu_b]}L
  \sim \frac 1{2\pi} \int_{-\pi}^\pi \mathrm{d}k\, \left.\frac{\mathrm d}{\mathrm d \mu_c}\right|_{\mu_c=0} \!\!\!\! \log \mathrm{Tr}\, \big[ e^{(\mu_c - \beta |D(k)|) c^\dagger_k c^{}_k} \big]
\]
and similarly for  $n_b = N_b/L$. By the Pauli exclusion principle,
\[
  \mathrm{Tr}\, \big[ e^{(\mu_c - \beta |D(k)|) c^\dagger_k c^{}_k} \big] = 1 + e^{\mu_c - \beta |D(k)|},
\]
hence
\begin{equation}\label{eq:mean-occ-num}
  \langle n_c \rangle_\beta = \langle n_b \rangle_\beta \sim \frac 1{2\pi} \int_{-\pi}^\pi \mathrm{d}k\, \frac 1{1 + e^{\beta |D(k)|}}.
\end{equation}

\begin{figure}
\centering\includegraphics[width=80mm]{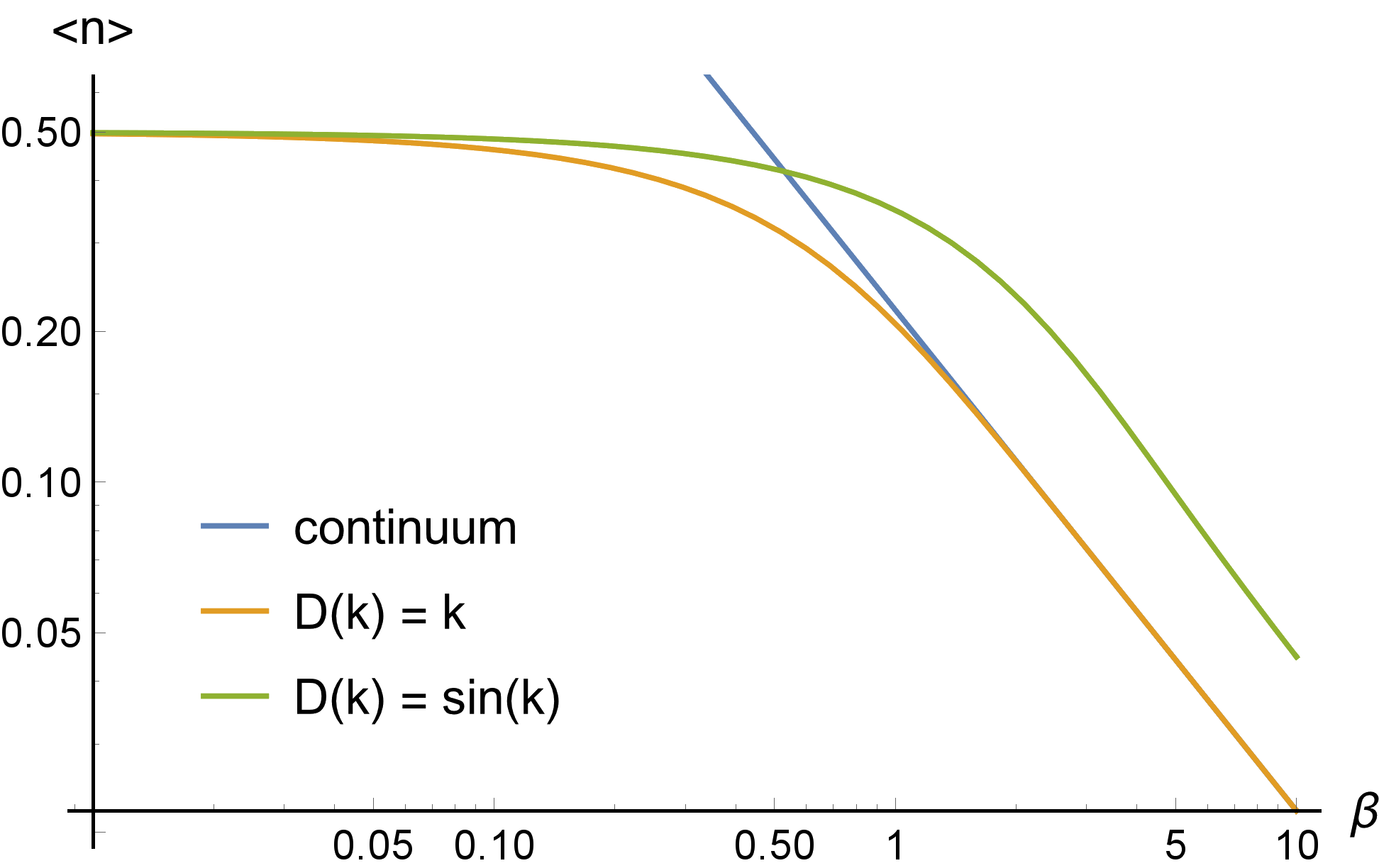}
\label{fig:fermion-density}
\caption{Mean fermion occupation number $\langle n \rangle_\beta$ at finite inverse temperature $\beta$, according to Eq.~(\ref{eq:mean-occ-num}). The naive lattice derivative (green) exhibits fermion doubling at low temperatures.}
\end{figure}

In the low temperature limit $\beta \to \infty$, this integral is highly susceptible to the behavior of $D(k)$ around its zeros. Indeed, for $D(k) = k$ we find that 
\[
  \langle n_{c/b} \rangle_{\beta} \sim \frac{\log 2}{\pi \beta} - \frac{\log [1 + e^{-\pi \beta}]}{\pi \beta}.
\]
The first term on the right hand side coincides with the continuum result, obtained by extending the range of integration in Eq.~(\ref{eq:mean-occ-num}) to infinity. The other term vanishes for $\beta \to \infty$, confirming that there is no fermion doubling. On the other hand, for $D(k) = \sin k$, we find
\[
  \langle n_{c/b} \rangle_{\beta \to \infty} \sim \frac{2\log 2}{\pi \beta},
\]
which is exactly twice as much as it should be. A comparative plot for general $\beta$ is given in Fig.~\ref{fig:fermion-density}.
%
\section{Renormalization of the $XY$ and Ising quantum chain}
\label{IsingSection}

The Hamiltonian of the $XY$ model on a chain with $N$ sites and periodic boundary conditions is defined by
\begin{equation}
\label{xyhamiltonian}
H = -\sum_{n=0}^{N-1} \Bigl[ \frac{1+\gamma}2 \sigma^x_{n}\sigma^x_{n+1} +\frac{1-\gamma}2 \sigma^y_{n}\sigma^y_{n+1} + h \sigma^z_n\Bigr] ,
\end{equation}
where $\gamma$ is the anisotropy parameter and $h$ the transverse field. In the special case $\gamma=1$ one obtains the Hamiltonian of the Ising quantum chain.

The Hamiltonian~(\ref{xyhamiltonian}) can be diagonalized following the standard methods by Lieb, Schulz, and Mattis~\cite{lieb1961two}. To this end one first rewrites the Hamiltonian in terms of the flip operators $\sigma_n^\pm = \frac12 (\sigma^x_n \pm \textrm{i} \sigma^y_n)$ by
\begin{equation}
H=h N \mathbbm 1 - \sum_{n=0}^{N-1} \Bigl[ \gamma\Bigl(\sigma^+_n\sigma^+_{n+1}+\sigma^-_n\sigma^-_{n+1}\Bigr) + \sigma^+_n\sigma^-_{n+1}+\sigma_n^-\sigma_{n+1}^++ + 2 h\sigma^+_n\sigma^-_n\Bigr].
\end{equation}
\paragraph{Jordan-Wigner transformation:}
The flip operators commute on different sites. In order to express the Hamiltonian in anticommuting operators, one performs a Jordan-Wigner transformation by introducing the operators
\begin{equation}
\psi_n:=(-\sigma^z_0)(-\sigma^z_1)\cdots(-\sigma^z_{n-1})\,\sigma^-_n\,,\qquad \psi_n^\dagger := [\psi_n]^\dagger.
\end{equation}
which obey the anticommutation relations $\{\psi_n,\psi_m\}=\{\psi^\dagger_n,\psi_m^\dagger\}=0, \;\; \{\psi_n,\psi_m^\dagger\}=\delta_{n,m}$. Rewriting the Hamiltonian in terms of these operator, the strings of $\sigma^z$-matrices drop out in all terms except for those between the last and the first site of the chain, where an operator $P=\prod_{n=0}^N (-\sigma^z_n)$ is encountered. Since this operator commutes with $H$ and has the eigenvalues $\pm 1$, the state space splits into two sectors. This means that the Hamiltonian can be written as
\begin{equation}
H=H_+\Pi_++H_-\Pi_-\,,
\end{equation}
where $\Pi_\pm=\frac12[\mathbbm 1 \pm P]$ are projectors on the respective sector. With $\psi=(\psi_0,\ldots,\psi_{N-1})$ and  $\psi^\dagger=(\psi^\dagger_0,\ldots,\psi^\dagger_{N-1})$ the sector Hamiltonians can be written in the bilinear form
\begin{equation}
\label{eq:xy-matrix}
  H_\pm = - \begin{bmatrix} \psi^{} \\ \psi^\dagger \end{bmatrix}^\dagger
  \begin{bmatrix} A^\pm & B^\pm \\ -B^\pm & -A^\pm \end{bmatrix}
  \begin{bmatrix} \psi^{} \\ \psi^\dagger \end{bmatrix}.
\end{equation}
where $A^\pm$ and $B^\pm$ are matrices of the form
\begin{equation}
  A^\pm = \frac 12
  \begin{bmatrix}
    2h &1 &  & &\mp 1 \\
    1 &2h &1 & &\\
    &\ddots &\ddots &\ddots &\\
    & &1 &2h &1 \\
    \mp 1 & & &1 &2h
  \end{bmatrix}
  \qquad \text{and} \qquad
  B^\pm = \frac{\gamma}2
  \begin{bmatrix}
    0 &1 &  & &\pm 1 \\
    -1 &0 &1 & &\\
    &\ddots &\ddots &\ddots &\\
    & &-1 &0 &1 \\
    \mp 1 & & &-1 &0
  \end{bmatrix}.
\end{equation}
\paragraph{Fourier transformation:}
As can be seen, the matrices $A^\pm$ and $B^\pm$ resemble a second and a first derivative on a lattice. It is therefore natural to Fourier-transform the Hamiltonian. Because of the corner matrix elements we expect two types of boundary conditions for the Fourier modes, namely, periodic b.c. in the negative sector and antiperiodic b.c. in the positive one. Assuming $N$ to be even we therefore introduce fermionic operators in momentum space
\begin{equation}
c_q = \frac{1}{\sqrt{N}} \sum_{n=0}^{N-1} e^{2\pi i n q/L } \,\psi_n
\end{equation}
and similarly $c^\dagger_q$, where the momentum index $q$ runs over the integers $-\frac N2,\ldots,+\frac N2-1$ in the negative sector and over half-integers $-\frac {N+1}2,\ldots,+\frac {N-1}2$ in the positive sector. The operators $c_q,c_q^\dagger$ still obey fermionic anticommutation relations. This allows the Hamiltonian to be rewritten as
\begin{equation}
\label{IsingC}
H_\pm\;=\; -{\sum_q}^\pm \Bigl( A^\pm_q \bigl(c^\dagger_q c_q-c_{-q}c^\dagger_{-q}\bigr) + B^\pm_q \bigl(c^\dagger_qc^\dagger_{-q}-c_{-q}c_q \bigr)\Bigr),
\end{equation}
where the sum $\sum^\pm_q$ runs over the ranges of $q$ specified above and where
\begin{equation}
A^\pm_q = h + \cos\frac{2 \pi q}{N}\,, \qquad B^\pm_q = i \gamma \sin\frac{2 \pi q}{N}
\end{equation}
are the eigenvalues of the two matrices $A^\pm$ and $B^\pm$ (with $\pm$ encoded in the range of $q$).
\paragraph{Decoupling of the edge modes:}
Now we would like to renormalize the Ising model by separating low and high momenta into two equal halves, as described in Sect.~\ref{sec:RGScheme}. In the positive sector, this is straight forward since the momentum index $q$ runs over $-\frac{N}{2}+\frac12,\ldots,-\frac32,-\frac12,\frac12,\frac32,\ldots,\frac{N}{2}-\frac12$, which splits naturally into an infrared part containing the modes $-\frac{N}{4}+\frac12,\ldots,\frac{N}{4}-\frac12$ and an ultraviolet part containing the other modes. However, in the negative sector, where $q$ runs from $-\frac N 2 $ to $\frac N 2-1$, the situation is more involved since the modes $q=\pm \frac N 4$ are exactly at the borderline between the IR and the UV sector. The corresponding contribution to the  Hamiltonian 
\begin{equation}
-\begin{bmatrix} c_{N/4} \\ c_{-N/4} \\ c^\dagger_{N/4} \\ c^\dagger_{-N/4}  \end{bmatrix}^\dagger
\begin{bmatrix} 
A^-_{N/4} &&& -B^-_{N/4} \\ 
& A^-_{N/4} & -B^-_{N/4} & \\
& B^-_{N/4} & -A^-_{N/4} & \\ 
B^-_{N/4} &&& -A^-_{N/4} \end{bmatrix}
\begin{bmatrix} c_{N/4} \\ c_{-N/4} \\ c^\dagger_{N/4} \\ c^\dagger_{-N/4}  \end{bmatrix}^\dagger
\end{equation}
couples the two edge modes via the off-diagonal elements. Therefore, we have to decouple the two edge modes in the negative sector \textit{before} the RG step is carried out. This can be done by means of a Bogoljubov transformation
\begin{equation}
c_{\pm N/4} \mapsto \tilde c_{\pm N/4} = \cos(\phi)\, c_{\pm N/4} \pm i \sin(\phi)\, c^\dagger_{\mp N/4}
\end{equation}
\textit{before} the momenta are separated. This transformation preserves the fermionic anticommutation relations and diagonalizes the contribution in the Hamiltonian by
\begin{equation}
\begin{bmatrix} c_{N/4} \\ c_{-N/4} \\ c^\dagger_{N/4} \\ c^\dagger_{-N/4}  \end{bmatrix}^\dagger
\begin{bmatrix} 
\Lambda_{N/4} &&&  \\ 
& \Lambda_{N/4} && \\
&& -\Lambda_{N/4} & \\ 
&&& -\Lambda_{N/4} \end{bmatrix}
\begin{bmatrix} c_{N/4} \\ c_{-N/4} \\ c^\dagger_{N/4} \\ c^\dagger_{-N/4}  \end{bmatrix}^\dagger
\end{equation}
where $\Lambda_{N/2}=\sqrt{(A^-_{N/4})^2-(B^-_{N/4})^2}$ is the fermionic excitation energy of the edge modes. Now it is possible to separate the modes and to assign them to the IR and the UV sector.

\paragraph{Renormalization and diagonalization:}
After separating the momenta, we diagonalize the Fourier-transformed Hamiltonian by Bogoljubov transformations in all non-diagonal excitation modes, considering the IR part as the new renormalized Hamiltonian. Following standard techniques, we arrive at the result that the renormalized Hamiltonian on a chain of $N$ sites, which has been obtained by $r$ renormalization steps from the original chain with $2^rN$ sites, can be written in the diagonal form
\begin{equation}
H_\pm^{(r)} = {\sum_q}^\pm \Lambda_q^{(r)} \eta_q^\dagger \eta_q 
\end{equation}
with fermionic operators $\eta_q^\dagger,\eta_q $ and the excitation energies
\begin{equation}
\Lambda_q^{(r)} = \sqrt{\bigl(h+\cos\frac{2\pi q}{2^r N} \bigr)^2 -\gamma^2 \sin^2 \frac{2\pi q }{2^r N} }
\end{equation}
In the case of the \textbf{critical} Ising model ($h=-1, \gamma=1$) reduce to 
\begin{equation}
\Lambda_q^{(r)} = \sqrt{2-2\cos\frac{2\pi q}{2^r N} }
\end{equation}
with the fixed point $\Lambda^*_q = \frac{2\pi q}{2^r N} $.

\paragraph{Conformal towers:}
%
\begin{figure}
\centering\includegraphics[width=95mm]{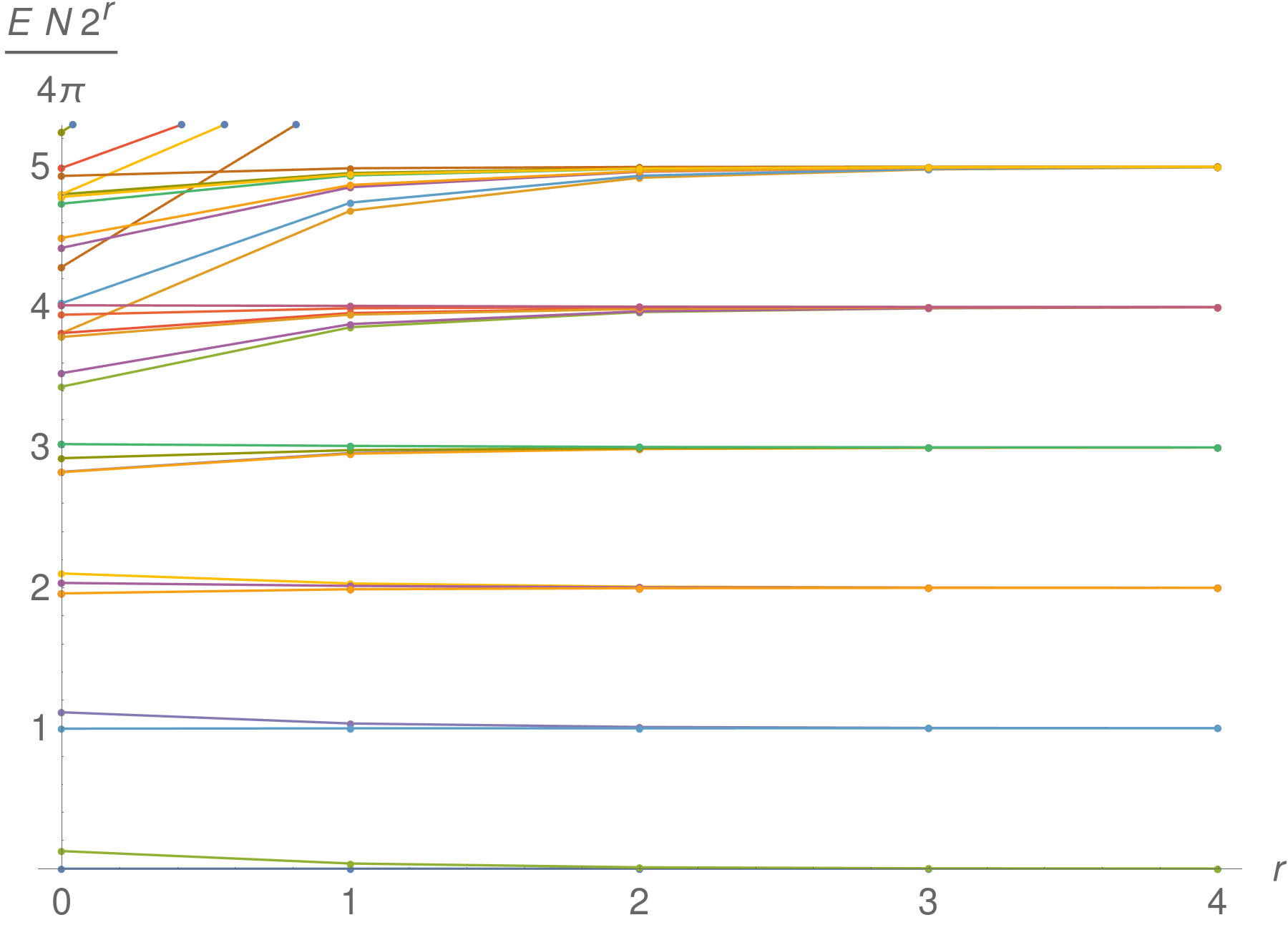}
\vspace{-2mm}
\caption{Emergence of ``conformal towers'' in the spectrum of the quantum Ising model under renormalization. The figure shows the rescaled energy eigenvalues of a chain with $N$ obtained by $r$ renormalization steps from an ordinary Ising model with $2^r N$ sites at criticality. The points of the spectra are connected by straight lines.}
\label{fig:towers}
\end{figure}
%
The linearity of $\Lambda^*_q$ implies that the spectrum  of the critical Ising Hamiltonian at the RG fixed point consists of equally-spaced levels. It is well-known that the low-lying spectrum of the Ising model exhibits such an equally-spaced structure, so-called \textit{conformal towers}, in the thermodynamic limit $N \to \infty$, and that this structure reflects the underlying conformal symmetry (Verma module) on large scales. It is remarkable that in the present study these conformal towers'' can also be observed in the RG-limit of \textit{finite} systems. This is demonstrated in Fig.~\ref{fig:towers}, where we plotted the rescaled spectrum against the number of renormalization steps. As can be seen, the spectrum becomes equidistant for large $r$. This means that the Ising model with long-range interactions at the fixed point exhibits a spectrum with more degeneracies and thus with a higher degree of symmetry which is a signature of the full conformal symmetry in the thermodynamic limit.

\section{Conclusions}

In this paper we have made  progress in the understanding of continuum quantum field theories on a lattice. Starting from the `bare' lattice action, we have defined an RG flow using infinitely extended wavelets, which ensures a perfect UV/IR decoupling. By construction, the fixed point of this flow exhibits a dispersion relation that coincides exactly with the continuum dispersion relation up to an energy scale given by the inverse lattice spacing. This implies that lattice effects are absent up to this energy scale. However, this comes at the price of introducing infinitely extended wavelets, which leads to highly non-local algebraically decaying interactions.

As applications, we have considered massive free fermions and bosons in $1+1$ dimensions, as well as the Ising quantum chain. We have demonstrated the usefulness of our method by calculating correlation functions and heat capacities. In particular, for massless free fermions we have shown how the fermion doubling problem is eliminated by the non-local nature of the wavelets. 

In view of constructing holographic dual spaces,
we note that previous approaches involving  finite-range wavelets cannot be generalized directly to the infinite-range case: This is due to the fact that correlations functions between UV and IR modes vanish by construction in the approach taken here. In the finite range wavelet approaches, precisely these correlation functions were used for defining the AdS radius. 
However, as demonstrated in Fig.~\ref{fig:towers}, our approach provides an exact renormalization of the Verma modules with the correct degeneracies up to a certain energy scale given by the inverse lattice spacing. This exact realization of the original conformal symmetry in a certain energy range is expected to be useful for a new holographic approach of constructing AdS/CFT duals using wavelets.
In this respect, a promising future direction is to consider the lattice version of coset minimal models \cite{Antonov:1995wv,Falceto:1992bf}. These are conjectured to be the field-theory duals of higher spin theories in the gravitational bulk \cite{Gaberdiel:2010pz,Gaberdiel:2012uj}. 

\section{Acknowledgments}
IR acknowledges Perimeter Institute for hospitality during the final stage of this project. 

\appendix
\section{Relation to other wavelet transformations}
\label{AA}

As already mentioned in the main part of the paper, SLAC-type derivatives, such as the one in Eq.~(\ref{longrangelaplacian}) are fixed points of a non-local renormalization scheme. In this appendix, we shall elaborate on how this generalizes to other wavelet induced renormalization schemes. 

To this end, recall from Fig.~\ref{fig:wavelet}, that a wavelet transformation implements
coarse-graining in two steps, the first being a lowpass filter, the second being a downsampler. A
generic signal $f_n$ is therefore coarse-grained as
\begin{equation}
  f_n \to f^{\mathrm{IR}}_n \equiv \sum_m S^{\mathrm{IR}}_{2n-m} f^{}_m,
\end{equation}
where $S^{\mathrm{IR}}_n$ is the real valued kernel of the lowpass filter. Similarly, the action of
a translationally invariant operator $O$ is coarse-grained as
\begin{equation}
  \sum_m O_{n-m} f_m \to \sum_m O^{\mathrm{IR}}_{n-m} f^{\mathrm{IR}}_m, \qquad
  O^{\mathrm{IR}}_{n-m} \equiv \sum_{l,p} S^{\mathrm{IR}}_{2n-l} O^{}_{l-p} S^{\mathrm{IR}}_{2m-p},
\end{equation}
which simplifies to
\begin{equation}
  O^{\mathrm{IR}}_n = \sum_{m,l} S^{\mathrm{IR}}_{2n-m+l} S^{\mathrm{IR}}_l O^{}_m.
\end{equation}

A fixed point with scaling dimension $d$ of the renormalization scheme is therefore solution of the
eigenvector equation
\begin{equation} \label{eq:realSpaceFP}
  \sum_m \Big(\sum_l S^{\mathrm{IR}}_{2n-m+l} S^{\mathrm{IR}}_l\Big) O_m = 2^{-d} O_n,
\end{equation}
which can be conveniently rewritten in momentum space as
\begin{equation} \label{eq:momentumSpaceFP}
  |S^{\mathrm{IR}}(k/2)|^2 O(k/2) + |S^{\mathrm{IR}}(k/2 + \pi)|^2 O(k/2 + \pi)
  = 2^{1-d} O(k).
\end{equation}

For the Shannon wavelet, characterized by
\begin{equation}
  S^{\mathrm{IR}}(k) = \sqrt{2} \theta(\pi/2 - |k|),
\end{equation}
we can easily see that the solutions of Eq.~(\ref{eq:momentumSpaceFP}) are given by monomials
$O(k) = k^d$, which corresponds to the SLAC-derivatives. All of these are non-local in real space,
since $k^d$ is not an analytic function of $e^{ik}$. In the other hand, the naively discretized
derivatives
\begin{equation}
  (\partial_{-1},\partial_0,\partial_1) = (-1/2,0,1/2), \qquad (\Delta_{-1},\Delta_0,\Delta_1) = (1,-2,1)
\end{equation}
with Fourier transforms
\begin{equation}
  \partial(k) = i \sin k, \qquad \Delta(k) = 2 \cos k - 2
\end{equation}
solve Eq.~(\ref{eq:momentumSpaceFP}) for the Haar wavelet lowpass filter
\begin{equation}
  S^{\mathrm{IR}}(k) = \frac 1{\sqrt{2}} \big( 1+e^{ik} \big).
\end{equation}
In this case, however, $\Delta$ turns out to have scaling dimension $1$ instead of $2$, which can be
traced back to the discontinuity of the Haar wavelet and thus its failure to properly renormalize
derivatives. 

\begin{figure}
\centering\includegraphics[width=80mm]{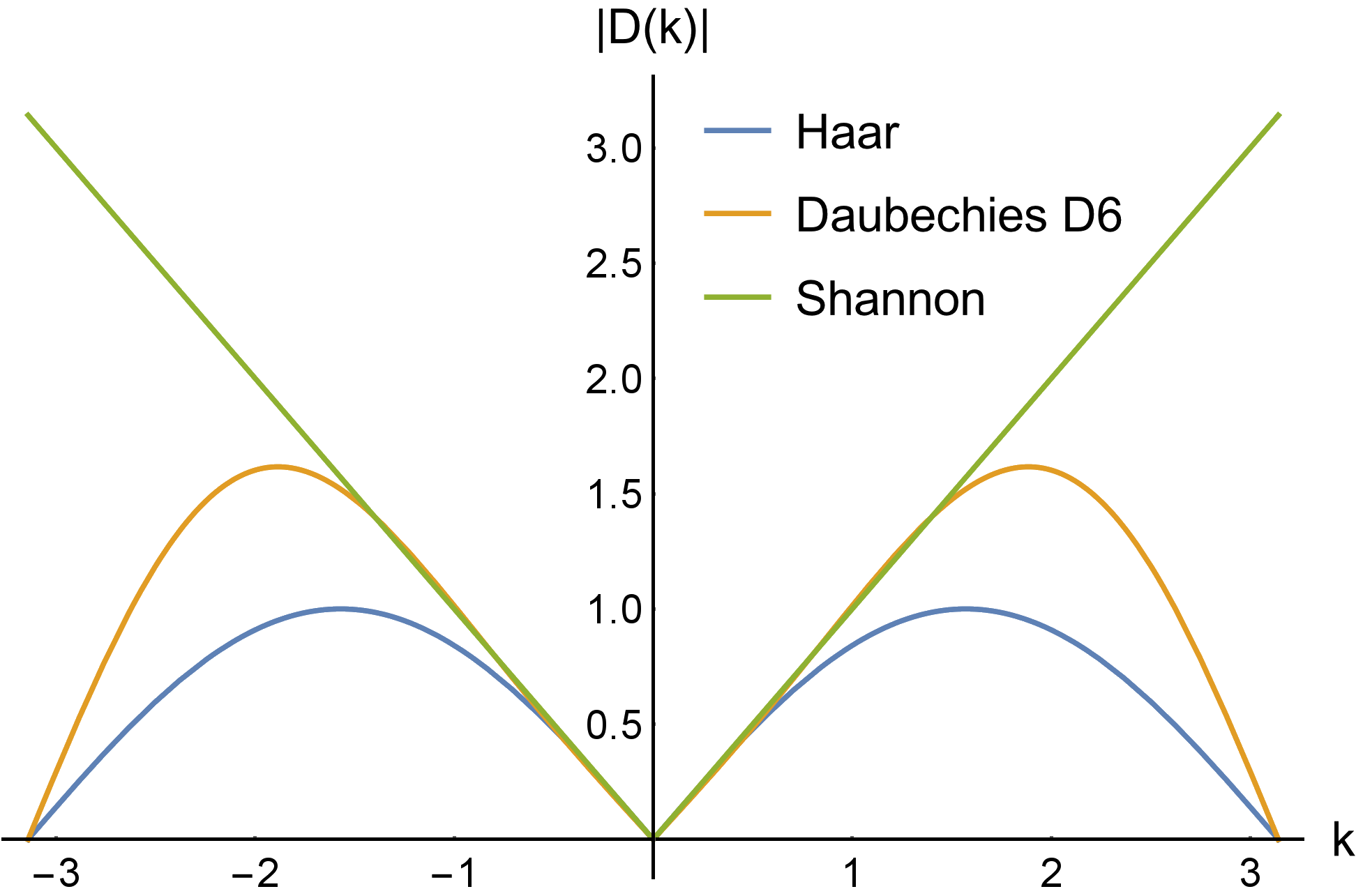}
\label{fig:dispersion-comparison}
\caption{Comparison of the gradient fixed points for different wavelet RG schemes. Going from Haar (blue) over Daubechies D6 (yellow) to the Shannon wavelet (green), we see that the problem of fermion doubling at $k=\pm \pi$ is resolved by an increase of the Fermi velocity $\partial_k |D(k)|$ for the doublers.}
\end{figure}

Intermediate steps between these two cases are given, e.g., by the Daubechies family of wavelets
and, as a last example, we will consider the Daubechies D6 wavelet, which is the first in this
family to properly renormalize first and second derivatives. Its lowpass filter is given by
\begin{equation}
  S^{\mathrm{IR}}(k) = \frac 1{16\sqrt 2} \big(1 + e^{ik}\big)^3 \Big(\big(1 + e^{ik}\big)^2
  + \sqrt{10}\big(1 - e^{ik}\big)^2 + \sqrt{5+2\sqrt{10}} \big(1 - e^{2ik}\big) \Big).
\end{equation}
The solutions of Eq.~(\ref{eq:momentumSpaceFP}) are found with basic linear
algebra. The ones with correct scaling dimension are
\begin{align*}
  O^{(0)}(k) &= 1 \\
  O^{(1)}(k) &= \frac{544}{365} i\sin k -\frac{106}{365} i\sin 2k + \frac{32}{1095} i\sin 3k
               +\frac 1{1460} i\sin 4k = ik + \mathcal O(k^7) \\
  O^{(2)}(k) &= -\frac{295}{56} + \frac{712}{105} \cos k - \frac{184}{105} \cos 2k
               + \frac 8{35} \cos 3k + \frac 3{280} \cos 4k = -k^2 + \mathcal O(k^6) \\
  O^{(3)}(k) &= -\frac {76}{25} i\sin k + \frac{179}{100} i\sin 2k - \frac 4{25} i\sin 3k
               -\frac 3{200} i\sin 4k = -i k^3 + \mathcal O(k^7) \\
  O^{(4)}(k) &= -\frac{81}{16} + \frac{41}5 \cos k -\frac{77}{20} \cos 2k + \frac 35 \cos 3k
               + \frac 9{80} \cos 4k = k^4 + \mathcal O(k^6) \\
  O^{(5)}(k) &= \frac{44}{13} i\sin k - \frac{31}{13} i\sin 2k + \frac 4{13} i\sin 3k
               + \frac 3{26} i\sin 4k = i k^5 + \mathcal O(k^7)
\end{align*}
A comparative plot of the above examples is given in Fig.~\ref{fig:dispersion-comparison}. For higher order
Daubechies wavelets, the fixed points lose more and more of their locality, converging to the SLAC
derivatives.
%
%


\newpage
\bibliographystyle{unsrt}
\bibliography{paper}
\end{document}